\documentclass[fleqn,usenatbib]{mnras}

\usepackage[T1]{fontenc}
\usepackage{ae,aecompl}

\emergencystretch=\maxdimen
\usepackage{graphicx}	
\usepackage{amsmath}	

\usepackage{newtxtext,newtxmath}

\newcommand{\dm}{{\rm pc \, cm^{-3}}} 
\def\arcsec{\hbox{$^{\prime\prime}$}}

\title[GP like emission from XTE J1810$-$197]{Radio and X-ray observations of giant pulses from XTE J1810$-$197}

\author[M. Caleb et al.]{M. Caleb,$^{1}$\thanks{E-mail: manishacaleb@gmail.com},
K.~Rajwade$^{1}$\thanks{E-mail: rkaustubh10@gmail.com},
G.~Desvignes$^{2,3}$,
B.~W.~Stappers$^{1}$,
A.~G.~Lyne$^{1}$,
\newauthor 
P.~Weltevrede$^{1}$,
M.~Kramer$^{2}$,
L.~Levin$^{1}$,
M.~Surnis$^{1}$
\\ \\  
$^{1}$Jodrell Bank Centre for Astrophysics, Department of Physics and Astronomy, The University of Manchester, Manchester, M13 9PL, UK\\
$^{2}$Max-Planck-Institut f\"ur Radioastronomie, Auf dem H\"ugel 69, D-53121 Bonn, Germany \\
$^{3}$LESIA, Observatoire de Paris, Universit\'e PSL, CNRS, Sorbonne Universit\'e, University Paris Diderot, Sorbonne Paris Cit\'e, 5 place Jules Janssen, 92195 Meudon, France\\
$^{4}$ Astrophysics Science Division, NASA Goddard Space Flight Center, Greenbelt, MD 20771, USA\\
}

\date{Accepted XXX. Received YYY; in original form ZZZ}

\pubyear{2021}

\begin{document}
\label{firstpage}
\pagerange{\pageref{firstpage}--\pageref{lastpage}}
\maketitle

\begin{abstract}
We present the results of two years of radio and X-ray monitoring of the magnetar XTE J1810$-$197 since the radio re-activation in late 2018. Single pulse analysis of radio observations from the Lovell and MkII telescopes at 1564 MHz and the Effelsberg telescope at 6 GHz has resulted in the detection of a total of 91 giant pulses (GPs) between MJDs 58858 and 59117. These GPs appear to be confined to two specific phase ranges ($0.473 \leq \phi\leq 0.502$ and $0.541 \leq \phi \leq 0.567$). 
We also observe that the first detection of GP emission corresponds to a minimum in the spin-down rate. Simultaneous radio and X-ray observations were performed on MJDs 59009 and 59096. The 0.5--10~keV X-ray spectrum from NICER is well characterised by a two component blackbody model that can be interpreted as two hot spots on the polar cap of the neutron star. The blackbody temperature decreases with time, consistent with the previous outburst, while the change in the pulsed fraction does not follow the same trend as was seen in the previous outburst. The radio and X-ray flux of XTE~J1810$-$197 are correlated during the initial phase of the outburst (MJD 58450-- MJD 58550) and an increase in the radio flux is observed later that may be correlated to the onset of GPs. We argue that the disparity in the evolution of the current outburst compared to the previous one can be attributed to a change in geometry of the neutron star.  
\end{abstract}




\section{Introduction}

Magnetars are an extreme class of young, slowly rotating ($P \approx 1-12$ s) neutron stars that possess high surface magnetic fields ($B \approx 10^{13}-10^{15}$ G) \citep[e.g.][]{KaspiBeloborodov}. They exhibit striking variability in their X-ray and soft $\gamma$-ray energies, powered by their decaying magnetic fields \citep{DT92}. This variability often manifests itself as common, millisecond-duration bursts, to rarer, bright month-long outbursts, which comprise many of the shorter bursts and a long tail or afterglow \citep[e.g.][]{KaspiBeloborodov}. Of the 25 confirmed magnetars\footnote{\url{http://www.physics.mcgill.ca/~pulsar/magnetar/main.html}}, only 5 have been detected to pulse in the radio \citep{CRH+06, CRH+07, LBB+10, EFK+13, CCC+20}. The radio single pulses exhibit extraordinary spectral and temporal phenomenology \citep{CRP+07, Kramer1810, Serylak,  DEK+18}, and have recently been suggested as possible progenitors to the enigmatic fast radio bursts \citep[FRBs;][]{MSF+20, PMP+20}.

The magnetar XTE J1810$-$197 was serendipitously discovered in 2003 by the Rossi X-Ray Timing Explorer \citep[RXTE;][]{IMS+04}. Subsequent X-ray follow-up of the source revealed a 5.75 second period, confirming it as an energetic neutron star with a measured magnetic field of 2.1$\times$10$^{14}$ G. It was later found to be associated with a radio source in the VLA MAGPIS survey at 1.4 GHz in 2005 \citep{HGB+05}. Observations by \cite{CRH+06}, resulted in the first detection of pulsed radio emission from a magnetar with a dispersion measure (DM) of 178 pc cm$^{-3}$. Since then, the magnetar has been extensively monitored by both ground and space-based telescopes \citep[e.g.][]{CRP+07, Pintore, Gotthelf, Torne}. Investigations of the radio polarization properties and flux densities has indicated dramatic variations in the polarization position angle over time, and a flat spectral index \citep{Kramer1810, DLB+19}. The average radio flux density and X-ray flux decreased in the 10 months following the discovery and remained stable for the next $\sim2$ years before eventually disappearing towards the end of 2008 \citep{CRH+16}. The magnetar was then in quiescence until the end of 2018 when it exhibited another large X-ray flare \citep{GHA+19} coinciding with the onset of radio emission \citep{LLS+18}. The source has been continuously monitored in the radio and X-ray since reactivation with strong pulsations detected at both wavelengths \citep{DEK+18, JMS+18, DLB+19, LGF+20}. Similar to the 2006 outburst \citep{CRH+06, Serylak}, the single pulses seen in the current outburst also exhibit `spiky' millisecond-width pulses in the radio \citep{DLB+19}. \cite{MJS+19} investigate this spiky emission and report on its similarities to those of the giant micropulses, thereby suggesting a possible link in the underlying emission mechanisms. The recent discovery of millisecond-duration radio pulses \citep{Andersen20, brb+20} from the magnetar SGR~1935$+$2154 in our own Galaxy coincident with high-energy emission in the X-ray and Gamma-rays \citep{MSF+20, zxl+20} indicates that at least some magnetars produce FRBs.

The Lovell and Mark II (MkII) telescopes at the Jodrell Bank Observatory (JBO) has been regularly monitoring the magnetar as part of a timing programme \citep{LLS+18}. Fortuitously, the magnetar has also been extensively monitored at soft X-ray wavelengths with the Neutron Star Interior Composition Explorer (NICER) instrument on board the International Space Station~\citep{Gendreau2016}.
In this paper we report on the long term timing and single pulse analysis of XTE J1810$-$197 with the JBO and Effelsberg telescopes and quasi-simultaneous X-ray analysis with the NICER instrument. Following the detection of giant pulse (GP) like emission in the radio, we performed simultaneous observations on 9 June 2020 and 4 September 2020 in the radio and X-rays with the Effelsberg and MkII telescopes, and the NICER telescope respectively. We present the results of the simultaneous observations and investigate the existence of quasi-periodicity in the radio single pulses.  In Section \ref{sec:obs} we present the radio and X-ray observations of XTE~J1810$-$197. The analysis and results of these observations are presented in Section \ref{sec:analysis}. We discuss possible implications of our findings in Section \ref{sec:R&D} and draw our conclusions in Section \ref{sec:conclusions}.

\section{Radio Observations and Data Reduction}
\label{sec:obs}

\subsection{Lovell and Mark II}

As detailed in \cite{LLD+19}, the magnetar XTE J1810$-$197 has been monitored since the beginning of 2009 using a combination of the 76-m Lovell telescope and the 26-m Mark II (MkII) telescope at JBO. 
The total intensity data for the Lovell and MkII telescopes were divided into sub-bands using a ROACH board \citep{BJK+16} and then channelised in real-time using the \textsc{digifil} algorithm from the \textsc{dspsr} software package \citep{2011VanStraten}. The resulting filterbanks were recorded with the Apollo backend instrument at a time resolution of 256 $\upmu$s covering a total bandwidth of 336 MHz centered at 1564 MHz. The data are divided into 672 frequency channels during acquisition. Each of the Lovell observations varied from 10 minutes to 30 minutes in duration, while the MkII observations were between 10 minutes and 1 hour long. For all observations, single pulse archives were created offline using the \textsc{dspsr} software package with an ephemeris obtained from the joint timing analysis of all the JBO and Effelsberg data, and dedispersed to a DM of 178 $\dm$. Since our data were affected by radio frequency interference (RFI) we used a combination of the \texttt{clfd}\footnote{\url{https://github.com/v-morello/clfd}} package described in \cite{MBC+19} and the Inter-Quartile Range Mitigation (IQRM) algorithm\footnote{\url{https://gitlab.com/kmrajwade/iqrm\_apollo}} to flag the channels with the most prominent RFI. Additionally, the data were visually inspected and manually cleaned using PSRCHIVE\footnote{\url{http://psrchive.sourceforge.net/}}
tools. An integrated pulse profile was created using the cleaned pulses, and folded and de-dispersed using the same ephemeris and DM as the single pulses.

\subsection{Effelsberg}

The 100-m Effelsberg radio telescope has also been monitoring the source since its outburst in December 2018. The observations presented here were performed using the S45mm single pixel receiver tuned at a central frequency of 6 GHz with a system equivalent flux density of 25 Jy averaged across the 4 GHz band. 
The  data were recorded in full Stokes mode using two ROACH2 backends, splitting the band into 4096 frequency channels, and have a time resolution of 131 $\upmu$s. Similar to the Lovell and MkII data analysis, the Effelsberg \texttt{psrfits} format search-mode data were processed for single pulses using \textsc{dspsr} and the same ephemeris as used for the Lovell and MkII data processing. Initial RFI mitigation was performed using the \texttt{clfd} package and further visual inspection and manual cleaning was carried out later using standard PSRCHIVE pulsar packages. The Effelsberg observations reported in this paper range from 11 minutes to 1 hour in duration. 
Simultaneous observations were undertaken with the MkII and Effelsberg telescopes on 9 June 2020 and 4 September 2020.

\section{X-ray Observations and Data Reduction}

Since the outburst in 2018, the  NICER X-ray telescope has been monitoring XTE~J1810$-$197, providing a high cadence X-ray data set over a span of more than 1.5 years since the onset of the outburst. NICER provides fast-timing spectroscopy observations with a large effective area in soft X-rays (0.2-12 keV) as an external payload on the International Space Station \citep{Gendreau2016}. After the onset of another period of bright radio activity from the magnetar in early 2020, we submitted a Target of Opportunity (ToO) proposal to observe XTE~J1810$-$197 simultaneously at X-ray and radio wavelengths to investigate any correlation in the X-ray and the radio lightcurves. In total, we were awarded three observations, each spanning 2~ks on the source. For the first two observing sessions (6 and 9 of June 2020), we observed simultaneously with the 26-m MkII telescope at 1564~MHz. The observations on 9 June 2020 were also shadowed by 6.0~GHz observations with the 100-m Effelsberg telescope. The third observing session on 4 September 2020 was shadowed by the MkII telescope. We note that the X-ray observations did not overlap with the entire span of the radio observations as continuous observing with NICER is significantly constrained by the orbital period of the International Space Station on which NICER is mounted.
Along with the simultaneous observing, we also downloaded the publicly available data on the magnetar from March of 2019 from the HEASARC NICER data archive\footnote{\url{https://heasarc.gsfc.nasa.gov/db-perl/W3Browse/w3table.pl?tablehead=name\%3Dnicermastr\&Action=More+Options}}. This resulted in a large X-ray data set on XTE~J1810$-$197 with an approximately monthly cadence over the span of 1.5 years. We do note that there is a gap in the X-ray monitoring of the source from November 2019 until March 2020 as the source was at very low Sun angles (Sun constrained).

\begin{figure}
\centering
\includegraphics[width=3.8 in]{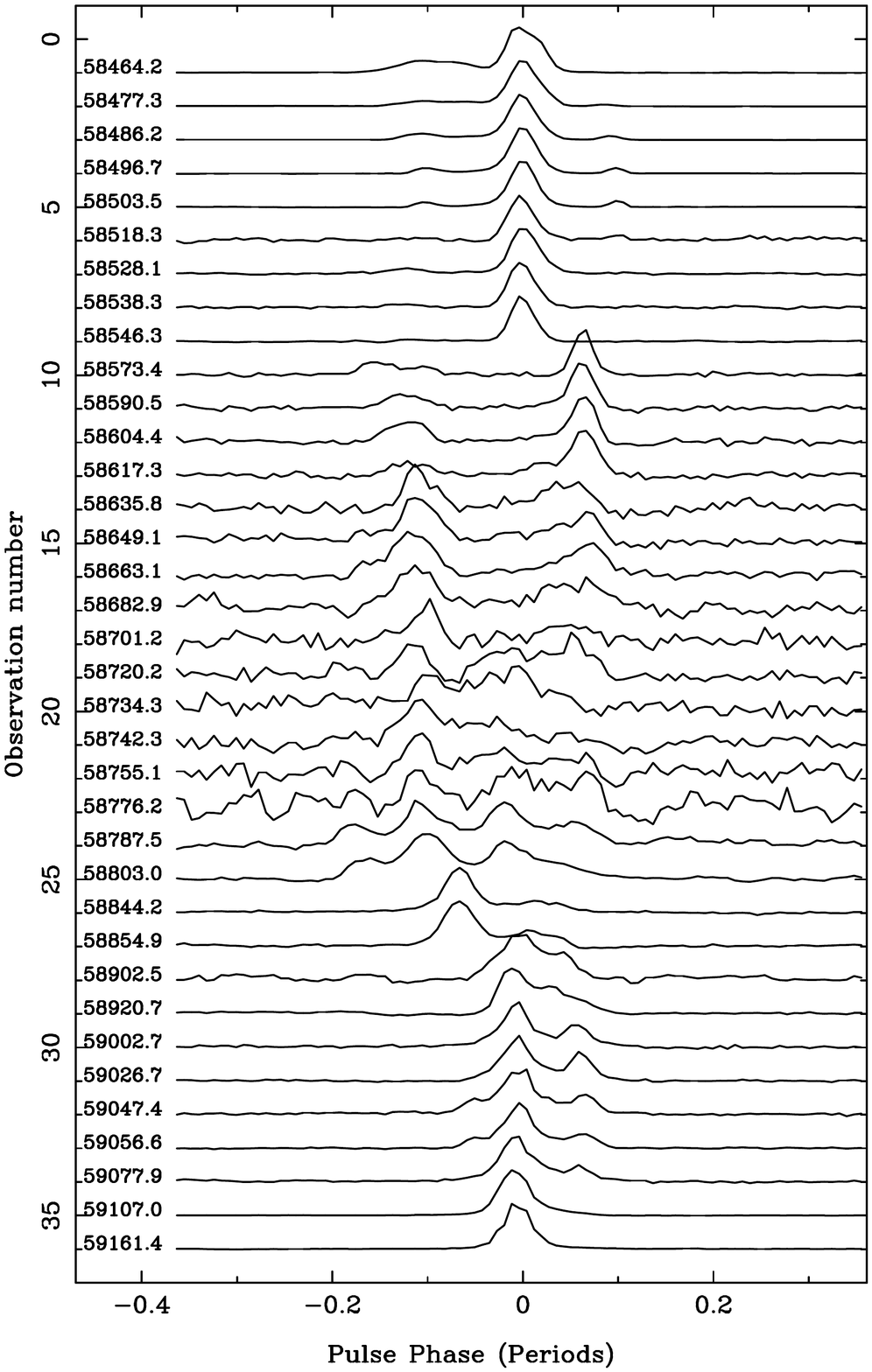}
\caption{Normalised total intensity profiles for observations
performed with the Lovell and MkII telescopes at 1564 MHz between 2018 and 2021. The profiles with the comparatively poor S/N between MJD~58590 and 58776 were carried out using the smaller MkII telescope (see Section \ref{sec:profevol}). Each profile has been averaged over 15-20 elapsed days. Phase zero corresponds to the position of the high-frequency component referred to in the text.}
\label{fig:profileEvol} 
\end{figure}

\begin{figure}
\includegraphics[width=3.6 in]{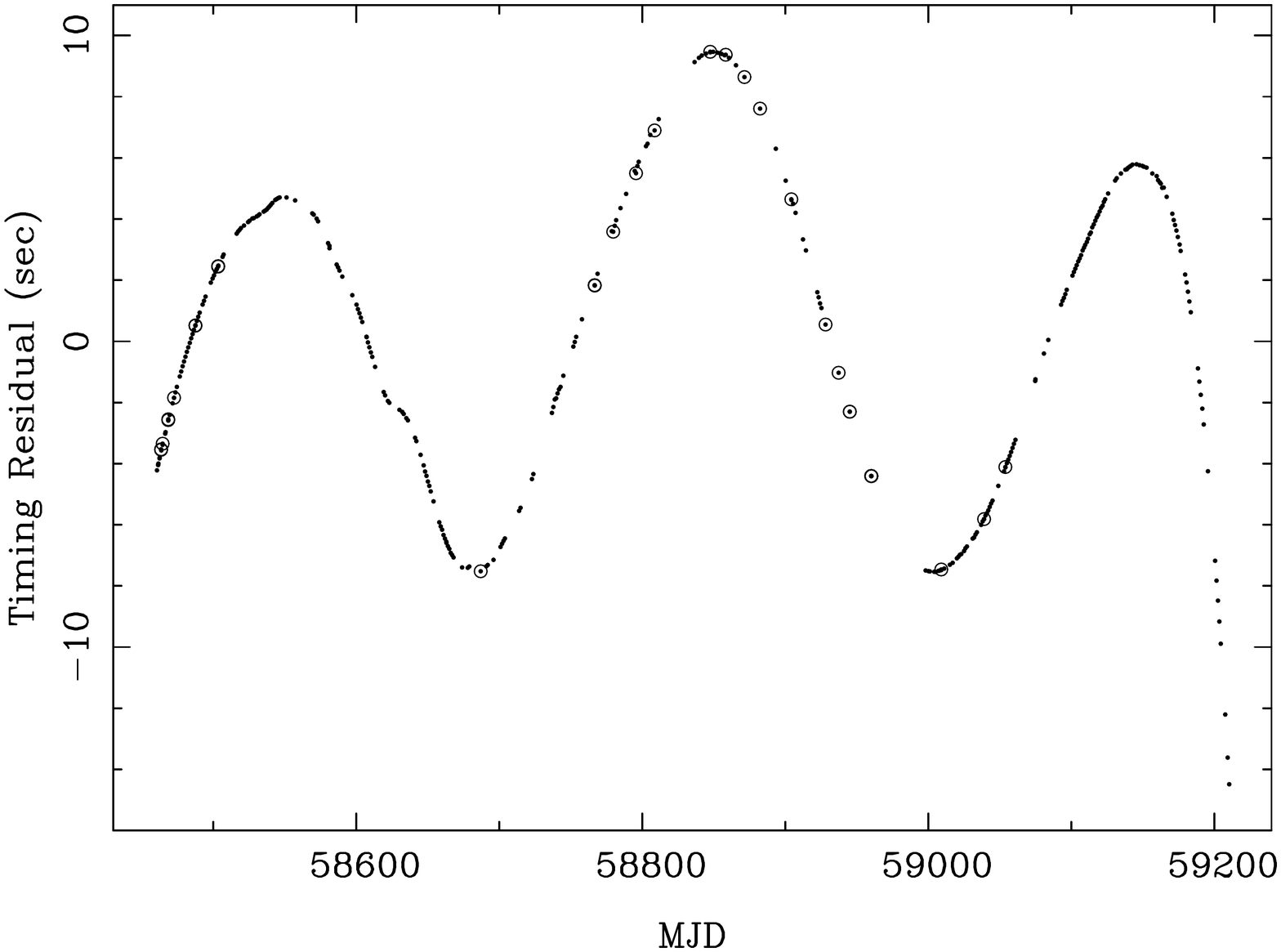}
\caption{Radio timing residuals for XTE~J1810$-$197 relative to the rotation model presented in Table \ref{tab:timingparams}. The small number of high-frequency data from Effelsberg are identified by surrounding open circles. The formal errors in the ToAs are less than the size of the symbols.}
\label{fig:residuals} 
\end{figure}

\begin{table}
\caption{Timing parameters of XTE J1810$-$197 for the data used in this work.}
\label{tab:timingparams}
\centering
\begin{tabular}{l l}
\hline \\ 
Parameter & Value\\ [1ex]
\hline \\
Right Ascension (J2000) [hh:mm:ss] & 18:09:51.140 \\
Declination (J2000) [dd:mm:ss] & $-$19:43:51.200 \\
DM [pc cm$^{-3}$] & 178 \\
Date range [MJD] & 58460.649106 -- 59210.427533 \\
Epoch of Frequency & 58848.5 \\
F0 [Hz] & 0.180445476(12) \\
F1 [Hz s$^{-1}$] & $-2.355(11) \times 10^{-13}$ \\
F2 [Hz s$^{-2}$] & $1.58(3) \times 10^{-20}$ \\
F3 [Hz s$^{-3}$] & $-8.09(14) \times 10^{-28}$ \\
F4 [Hz s$^{-4}$] & $-1.57(4) \times 10^{-34}$ \\
\hline
\end{tabular}
\end{table}

For calibration and filtering of the data, we used HEASOFT version 6.25 and NICERDAS version 2018-10-07 V005. We applied the standard filtering criteria (excluding events acquired during times of South Atlantic Anomaly passage and with pointing offsets greater than 54$\arcsec$; including data obtained with Earth elevation angles greater than 30$^{\circ}$ above the dark limb and 40$^{\circ}$ above the bright limb). The source was Sun constrained until February of 2019. The observations conducted during February 2019 were severely affected by optical loading of the X-ray detectors of NICER as the angle from the Sun was never above 60$^{\circ}$. We used the standard NICERDAS suite \textsc{nicerl2} to calibrate, filter and extract good time intervals (GTIs) for all our observations. For each GTI, we extracted the corresponding XTE~J1810$-$197 spectrum. For the 6, 9 June 2020 and 4 September 2020 observations, we also extracted the lightcurves for XTE~J1810$-$197 with a time bin size of 1 second. Our choice of a large bin size was to gather sufficient photon counts per pulse phase bin above the noise threshold in order to see any correlations with the radio data. Because NICER consists of non-imaging detectors, the NICER team has developed a space weather-based background model, which estimates spectral contributions from the time-dependent particle background, optical loading from the Sun, and diffuse sky background using a library of observations of  `blank sky' fields (K. C. Gendreau et al. 2021, in preparation). Using the \textsc{nibkgestimator} tool provided by the NICER team\footnote{\url{https://heasarc.gsfc.nasa.gov/docs/nicer/tools/nicer_bkg_est_tools.html}}, we estimated the background spectrum for each observation segment separately. The lack of any other bright X-ray source in the field of XTE~J1810$-$197 excludes the possibility of source confusion when reducing NICER data. 

\begin{figure}
\includegraphics[width=3.6 in]{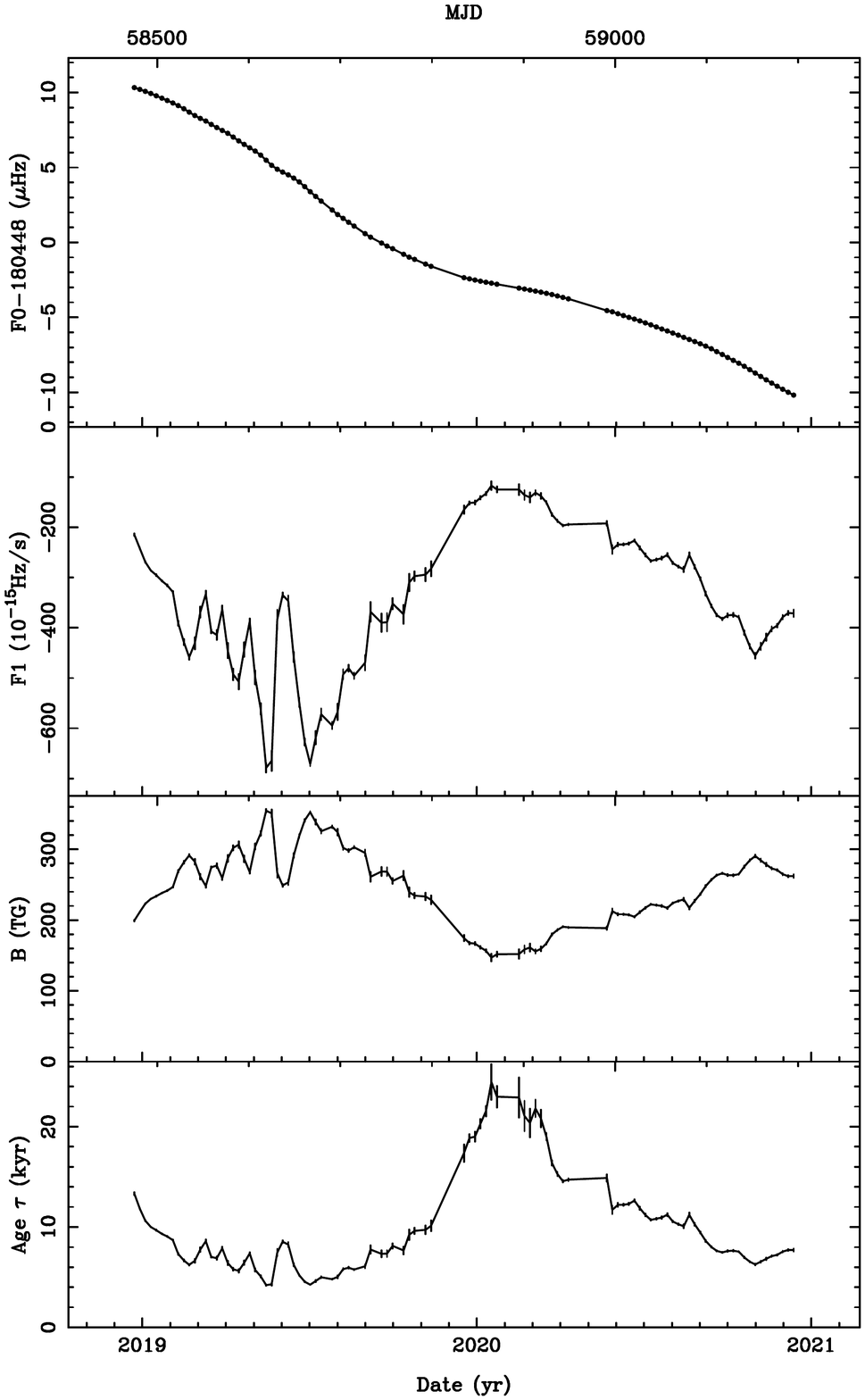}
\caption{Evolution of the spin frequency (F0), frequency derivative (F1), magnetic field (B $= 3.2 \times 10^{19}\,\sqrt{P\dot{P}}$) and characteristic age ($\tau = P/2\dot{P}$) as a function of time. Timing fits of F0 and F1 were made to the data over successive 30-day periods, starting at intervals of 10 days, so that adjacent points are not completely independent.}
\label{fig:timing} 
\end{figure}

\section{Radio Data Analysis and Results}
\label{sec:analysis}

\subsection{Profile evolution}
\label{sec:profevol}

The radio profile of XTE J1810$-$197 has undergone considerable changes since 2019 \cite{LLD+19}, exhibiting discrete components which change their relative flux densities with time and frequency. The total intensity profiles from the Lovell and MkII telescopes at 1564 MHz are shown in Figure \ref{fig:profileEvol}. Each profile is averaged over 15 - 20 days.
The main pulse profile peak has changed from being flanked by components on either side (MJD 58503), similar to the last few epochs in \cite{LLD+19}, to having a precursor component on the left, centred around phase -0.12 periods (MJD 58604). At MJD 58635, the main peak dropped in intensity while the precursor component got brighter. Between MJDs 58720 and 58803 inclusive, the profile underwent a number of changes, transitioning from a two- to a four- to a three-peaked profile. The comparatively lower signal-to-noise ratios (S/N) of these profiles are in part because they were obtained using the smaller MkII telescope, which has a sensitivity of about 1/10 of that of the Lovell, and in part because the pulsar flux density was low at that time (see Figure~\ref{fig:flux_evo}). From MJD 58920 the profile transitioned back to having two-components (but with a much temporally closer post-cursor rather than a precursor), and evolved into a stable single peak after MJD 59077. The smaller data set of higher-frequency observations carried out at Effelsberg also show drastic changes, but the profile components have differing spectra so that approximately contemporaneous Jodrell Bank profiles look quite different. The time alignment of the Jodrell Bank and higher-frequency Effelsberg profiles is discussed below.

\begin{figure}%
    \includegraphics[width=3.3 in]{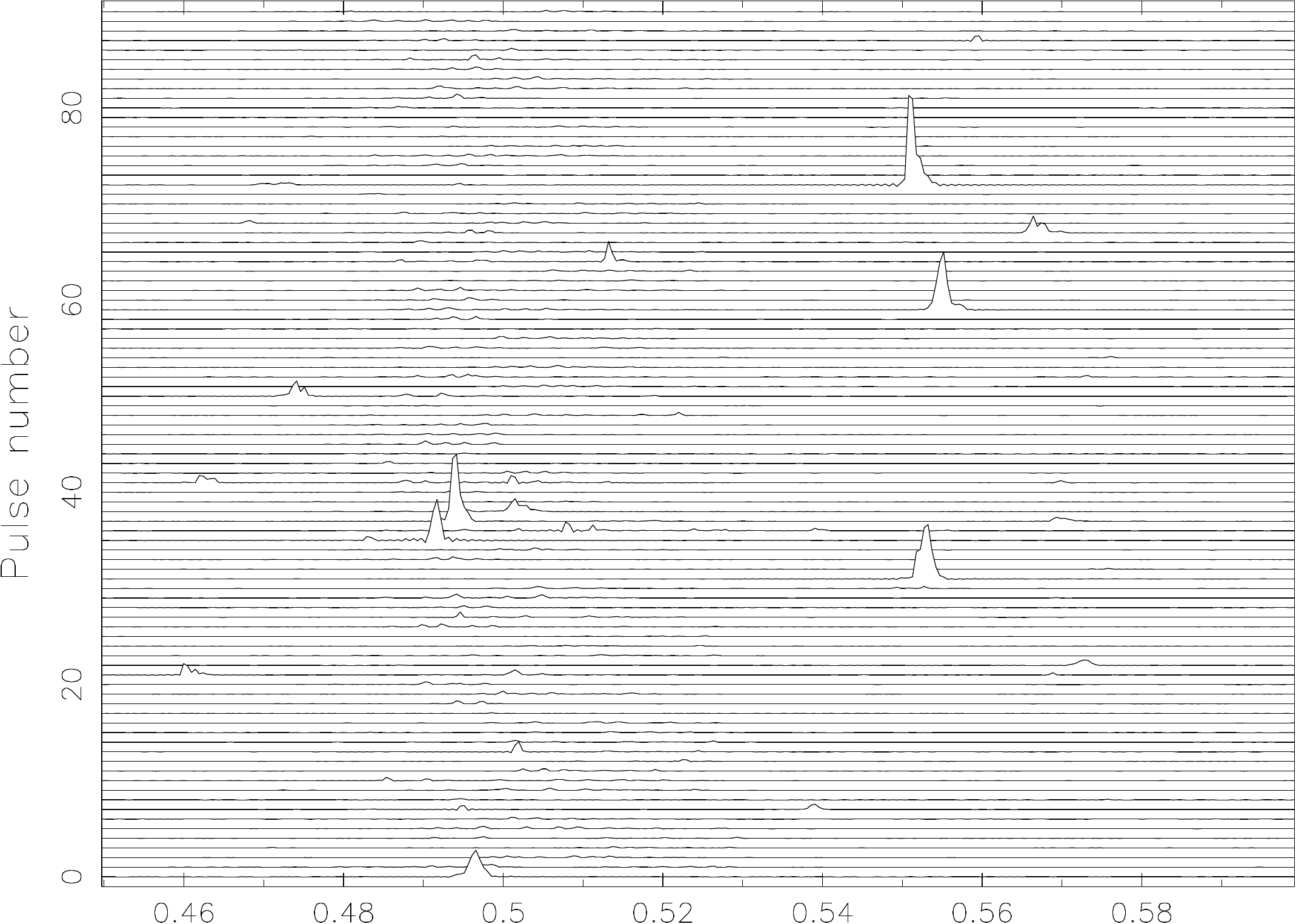} 
    \includegraphics[width=3.3 in]{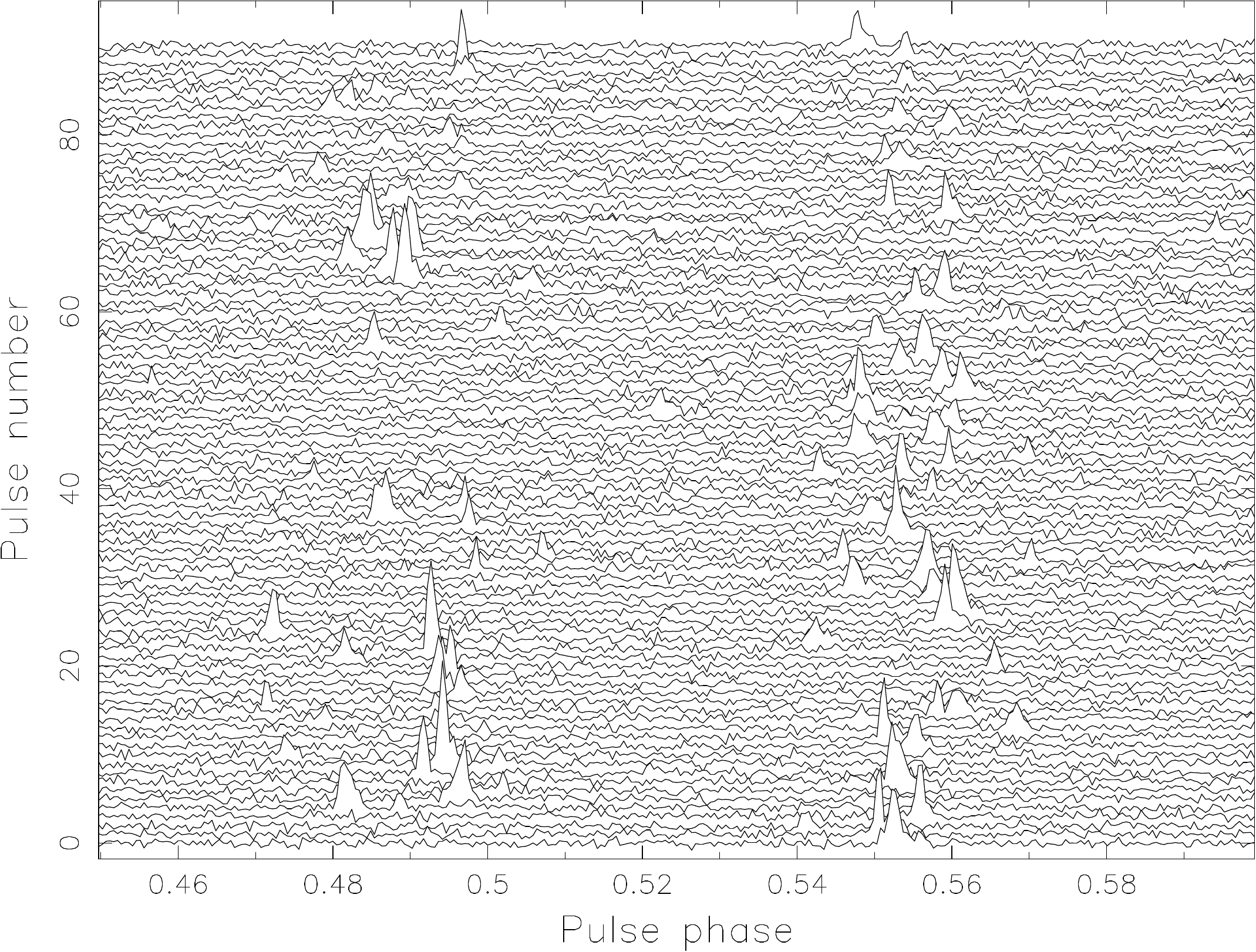}
    \caption{Pulse stack of XTE~J1810$-$197 single pulses from simultaneous Effelsberg and MkII observations on MJD 59009. \textit{Top panel}: Effelsberg single pulses observed at 6 GHz. \textit{Bottom panel}: MkII single pulses at 1564 MHz.}%
    \label{fig:stack}
\end{figure}

\subsection{Timing analysis}
\label{sec:timing}

A phase-coherent timing solution was obtained using a combination of the Lovell, MkII and Effelsberg data spanning $\sim2$ years from the start of the recent outburst in December 2018. Times of arrival (ToAs) were generated by cross-correlating the integrated profile for each MJD with a standard profile using standard PSRCHIVE pulsar timing tools. This process was more difficult than usual because of the profile changes described above and seen in Figure~\ref{fig:profileEvol}. For the Jodrell Bank data, multiple standard profiles were created by aligning and integrating observations from MJDs 58460 to 59210. Different standard profiles were adopted when significant changes in profile shape occurred. The changes mostly occurred abruptly, without any immediate change in phase drift rate across the event, relative to a pre-event ephemeris.  A new standard profile was adopted and adjusted in phase so as to cause no discontinuous step in timing residuals. This process was repeated at each of the four major changes, resulting in 5 standard profiles, which are all aligned with the phase of the intense dominant component present at around MJD~58500.

Inspection of the multi-frequency data set shows that this dominant component has a very different spectrum from other components and is even more dominant at higher frequencies, so that it is always present during the Effelsberg observations and nearly always the brightest feature at 6 GHz. ToAs for this feature were obtained using a standard profile which was typical for this component and which was aligned with the dominant component in the standard template for the Jodrell Bank data around MJD~58500. The ToAs were processed using the TEMPO2\footnote{\url{https://www.atnf.csiro.au/research/pulsar/tempo2/}} \citep{HEM+06} software package. In order to investigate the quality and consistency of the timing data, we fitted for the spin frequency (F0) and its first four derivatives, keeping the DM fixed at 178 pc cm$^{-3}$. We present the results Table~\ref{tab:timingparams}. The residuals of the ToAs relative to this model are shown in Figure~\ref{fig:residuals} and show only smooth variations in rotation. There is no indication of any discontinuities, which might have arisen at changes in standard profile. Additionally, the high-frequency Effelsberg points are satisfactorily consistent with the Jodrell Bank data. It should be noted that the variable nature of the profiles in this source means that the standard profiles are often imperfect matches to the observed profiles, which frequently vary in detail from hour-to-hour and day-to-day.  Close inspection of the timing residuals indicates a typical rms jitter of about 50 ms or 0.01 in pulse phase, increasing where the pulse is less distinct.

The timing model derived above suggests that the spin frequency derivative (F1) is changing rapidly, and we have performed a series of stride fits to the data to investigate the evolution of |F1| with time (Figure~\ref{fig:timing}). Indeed, the spin-down rate, |F1|, varies by a factor of about 5, reaching a minimum around the middle of these observations. The errors in the fitted values reflect the measurement jitter and are usually small compared with most of the structure seen in the figure.

\begin{figure*}
\centering
\includegraphics[width=7 in]{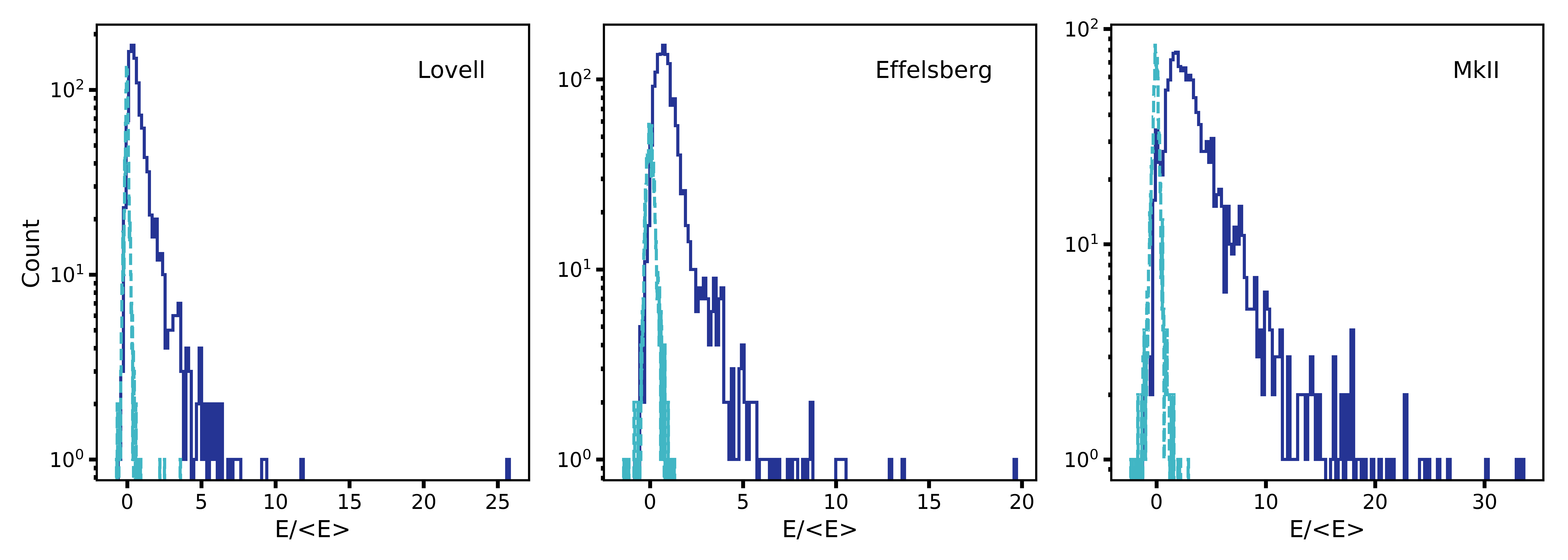}
\caption{Pulse energy distributions of the single pulses for the Lovell MJDs shown in Figure \ref{fig:GPfluxes} and listed in Section \ref{sec:PEDs}, MkII MJDs listed in Section \ref{sec:PEDs} and Effelsberg MJDs shown in Figure \ref{fig:LRCED}. The on-pulse and off-pulse distributions at each telescope are represented by the solid and dashed lines respectively. \textit{Left panel}: Single pulses observed at the Lovell telescope at 1564 MHz. \textit{Middle panel}: Single pulses observed at the Effelsberg telescope at 4 GHz. \textit{Right panel}: MkII single pulses at 1564 MHz.}
\label{fig:pulsedist} 
\end{figure*} 

\begin{figure}
\centering
\includegraphics[width=3.5 in]{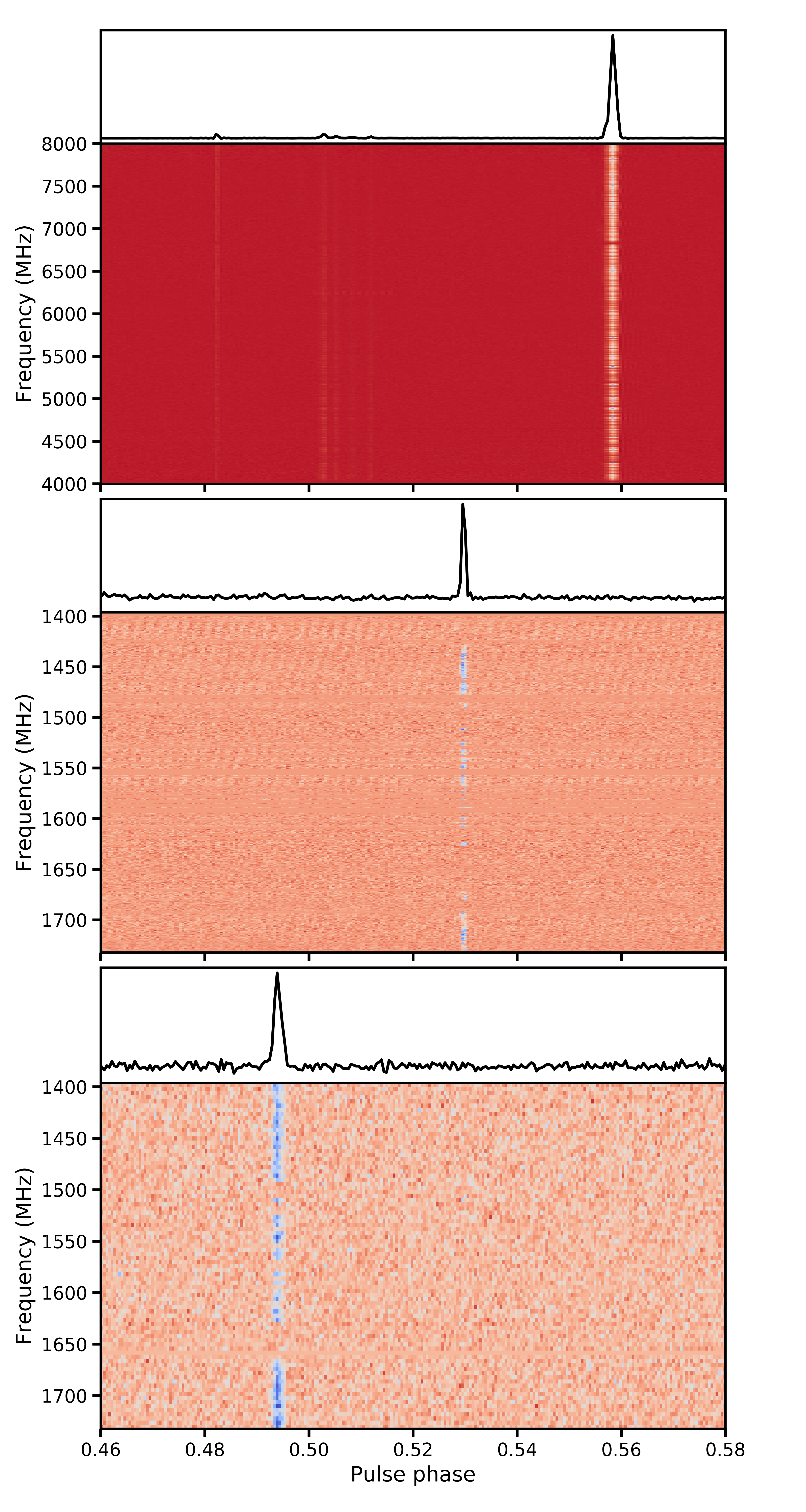}
\caption{Time-aligned dynamic spectra of example GPs detected with the Effelsberg (top panel; MJD 59009), Lovell (middle panel; MJD 58923) and MkII (bottom panel; left: MJD 59015) radio telescopes. The top panel in each pulse shows the frequency-averaged pulse profile. The data are uncalibrated, and the flux densities are in arbitrary units. The Effelsberg, Lovell and MkII pulses have been decimated to 512, 84 and 84 frequency channels and 2.7 ms resolution respectively. The background striations seen in the pulse in the middle and bottom panels are due to RFI and not intrinsic to the source. The top panel showcasing the Effelsberg GP also shows barely discernible normal emission between phases 0.48 and 0.52.}
\label{fig:GPspectra} 
\end{figure}

\subsection{Spiky giant pulse-like emission}
\label{sec:SpikeyEmission}

The strongest argument for the presence of a distinctive `spiky' emission is visible in an example pulse stack shown in Figure \ref{fig:stack}. Bright, narrow (see Table \ref{tab:GP_MJDs}, for values of S/Ns and widths) pulses visible across the whole observing band appear to dominate the emission suggesting that they may have different properties from the rest of the observed emission. Though the pulse stack is dominated by the spiky emission, we note that there is an almost indiscernible background of weak emission in the Effelsberg pulses. We refer to this as `normal' emission throughout the paper. The MkII pulses are however purely dominated by spiky emission and the absence of normal emission can be reconciled by the difference in sensitivities between the Effelsberg and MkII telescopes.
Similar to \cite{MJS+19}, we examine the possibility of some of the the spiky emission being \textit{true} giant pulses (GPs) by definition (see Sections \ref{sec:GPs} and \ref{sec:PEDs} for more details), by computing the pulse energy distributions. We note that throughout the paper, the Lovell and MkII data are flux calibrated, while we use the flux density values from the radiometer equation for the Effelsberg data.

\subsubsection{Pulse energy distributions}
\label{sec:PEDs}
We calculate the pulse-energy distributions of XTE~J1810$-$197 to compare the observed bright pulses with the GP phenomenon \citep[e.g.][]{Serylak}. To do this, we calculate the total single pulse energies and compare them with the average pulse energy. For each individual pulse, we calculated the average flux densities within the on-pulse region after subtracting the baseline noise using \textsc{psrsalsa}\footnote{\url{https://github.com/weltevrede/psrsalsa}} \citep{psrsalsa}. The off-pulse energy was determined in the same way using an equal number
of off-pulse bins. Visual inspection of the single pulse Effelsberg data showed a broad envelope of narrow, spiky emission or `normal emission' as well as GP like emission distinctly separated in phase from the normal emission. For the Effelsberg data, the single pulse energies containing both the normal and GP-like emission for each observation are compared with the average pulse energy across all the observations. As these data are not flux calibrated, we use the radiometer equation to convert the digitizer counts to Jansky. We note that there will be a slightly larger uncertainty in the flux due to possible changes in the system equivalent flux density across the different days. We also note that the bright, spiky emission cannot be attributed to interstellar scintillation as the scintillation timescale from the NE2001 model \citep{CL2003} is of the order two pulse periods. 

Visual inspection of the Lovell single pulse data from MJDs 58900, 58904, 58905, 58922, 58923, 58924 and 58925 showed negligible amounts of low-level normal emission. Most pulses are dominated by single bright peaks or multiple bright peaks. Similarly, the MkII single pulse data from MJDs 59000, 59001, 59009, 59015, 59020, 59021, 59022, 59023, 59036, 59053, 59054 and 59096 also exhibit only bright, narrow peaks with insignificant amounts of normal emission. 
The average pulse energy for each Lovell and MkII observation was calculated from the integrated profiles of only the underlying low-level normal emission obtained by clipping all the rotations that showed bright GP-like emission. 
Therefore in Figure \ref{fig:pulsedist}, the single pulse energies for each Lovell and MkII observation are compared with the corresponding average pulse energy for that observation rather than the overall average across all observations as was done for the Effelsberg data. While definitions of GPs vary, a common theme is that single pulse energies above a threshold of $10\langle E \rangle$ are defined as GPs and are significantly narrower than the average pulse window. \citep[e.g.][]{Cairns2004}. By this definition, we observe 7, 2 and 82 GPs in the Effelsberg, Lovell and MkII data, which correspond to rates of 1.7, 0.4 and 5.6 bursts hour$^{-1}$ respectively. We note that the MkII observations are non-contemporaneous with the Lovell observations, and there may have been an evolution in the burst rate at 1564~MHz over time. We emphasize that though several pulses are GP-like upon visual inspection, they do not meet the classic criterion of the total energy exceeding 10 times the average pulse energy. However, these pulses may indeed be drawn from the same class as the GPs. While \cite{MJS+19} also report on GP-like emission in their data, the mean flux densities of these bursts do not exceed the conventional threshold.

\begin{figure}
\centering
\includegraphics[width=3.3 in]{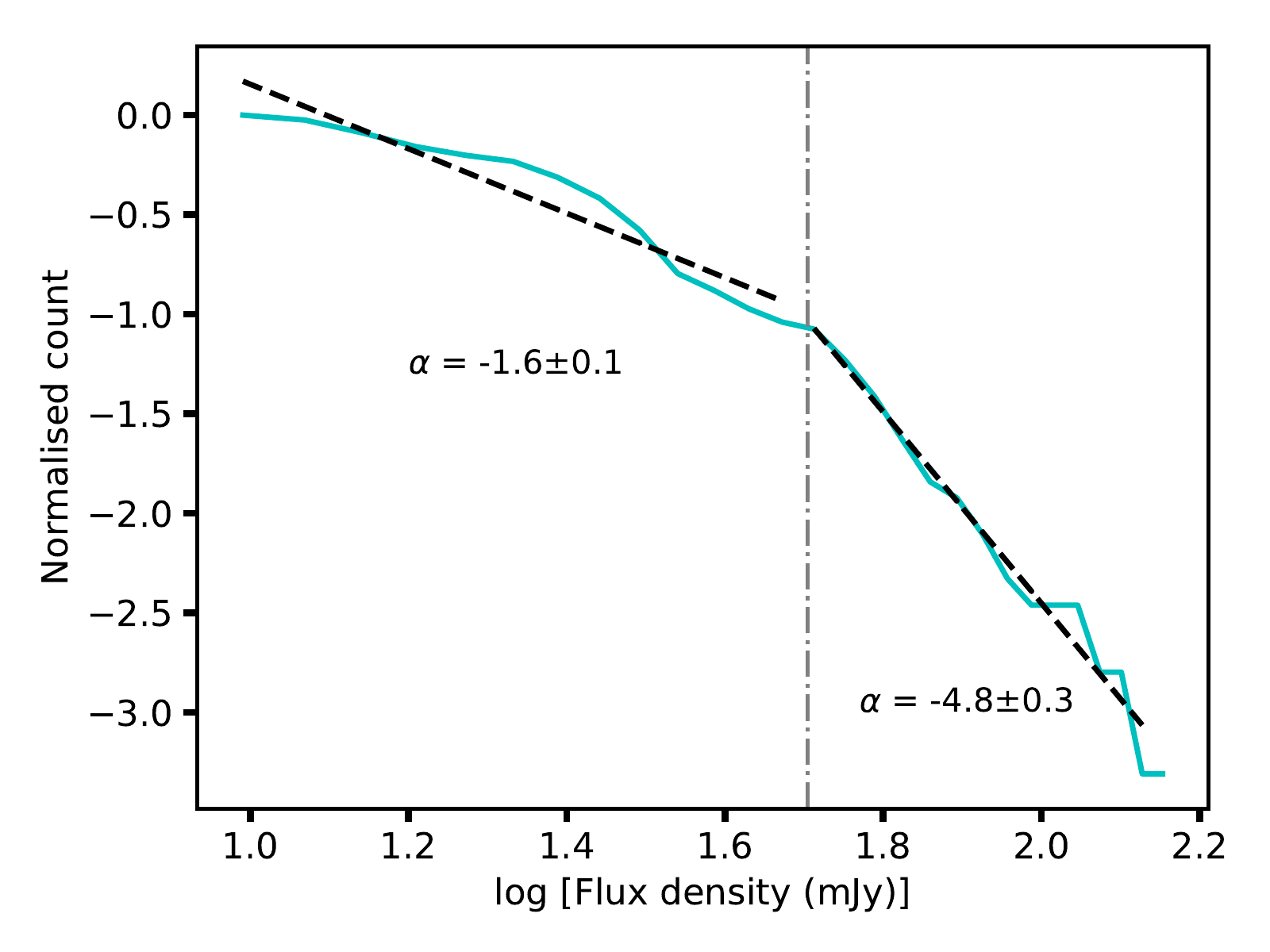}
\caption{A plot of the cumulative number of GPs detected with the MkII telescope versus their fluxes. The distribution is described by a broken power-law with a break at $5.6$ mJy and slopes of $\alpha = -1.6\pm0.1$ at the fainter end and $\alpha = -4.8\pm0.3$ at the brighter tail end. }
\label{fig:plfit} 
\end{figure} 

\begin{figure}
\centering
\includegraphics[width=3.3 in]{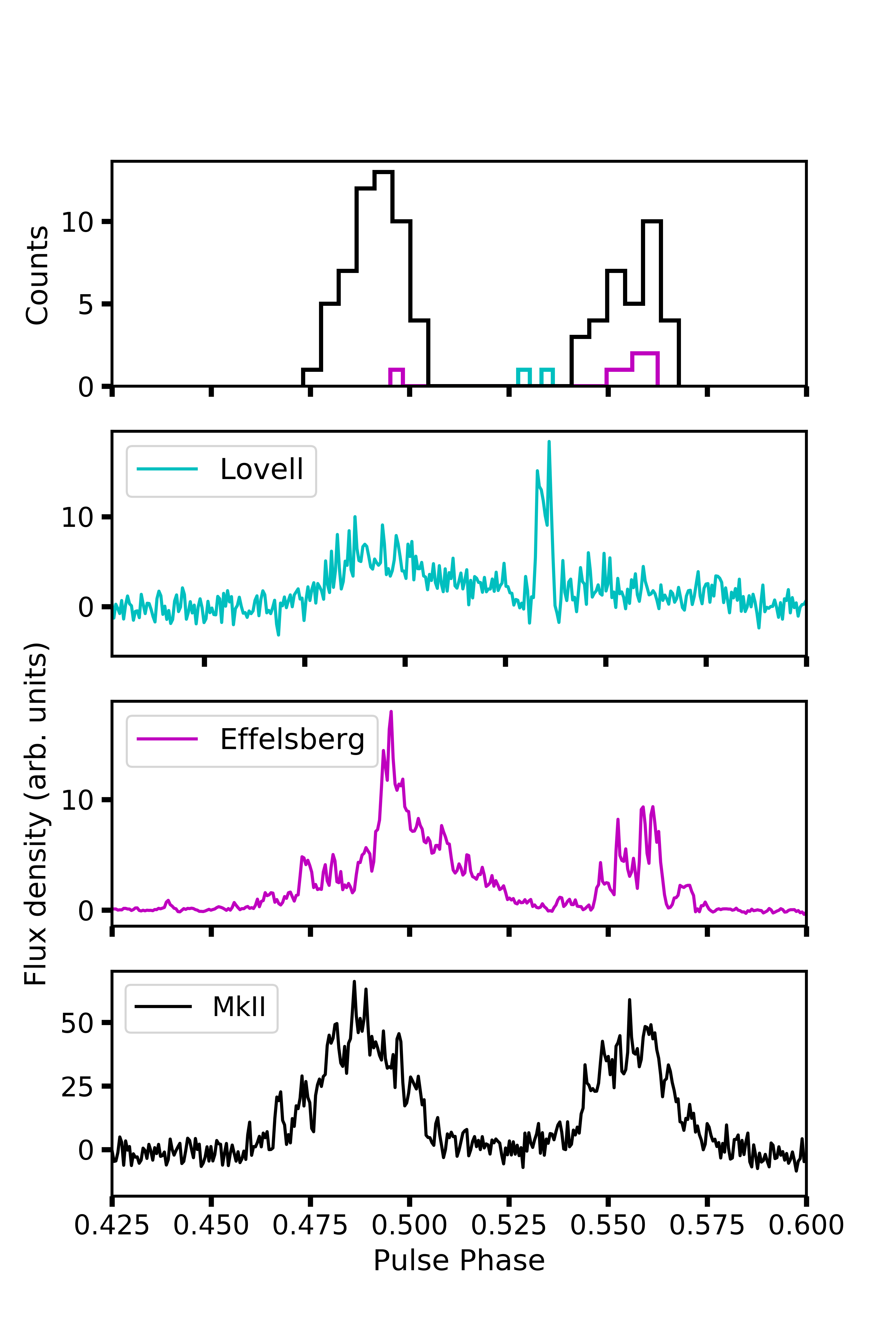}
\caption{Distribution of phases of the GPs detected at the Lovell, Effelsberg and MkII telescopes. The Lovell and Effelsberg pulse profiles are for MJDs 58923 and 59009, respectively. The MkII profile is the average across all MJDs. Similar to the Crab GPs, the GPs reported here also appear over a narrow range of pulse phases.}
\label{fig:phasedist} 
\end{figure} 
 
\begin{figure*}
\includegraphics[width=7.2 in]{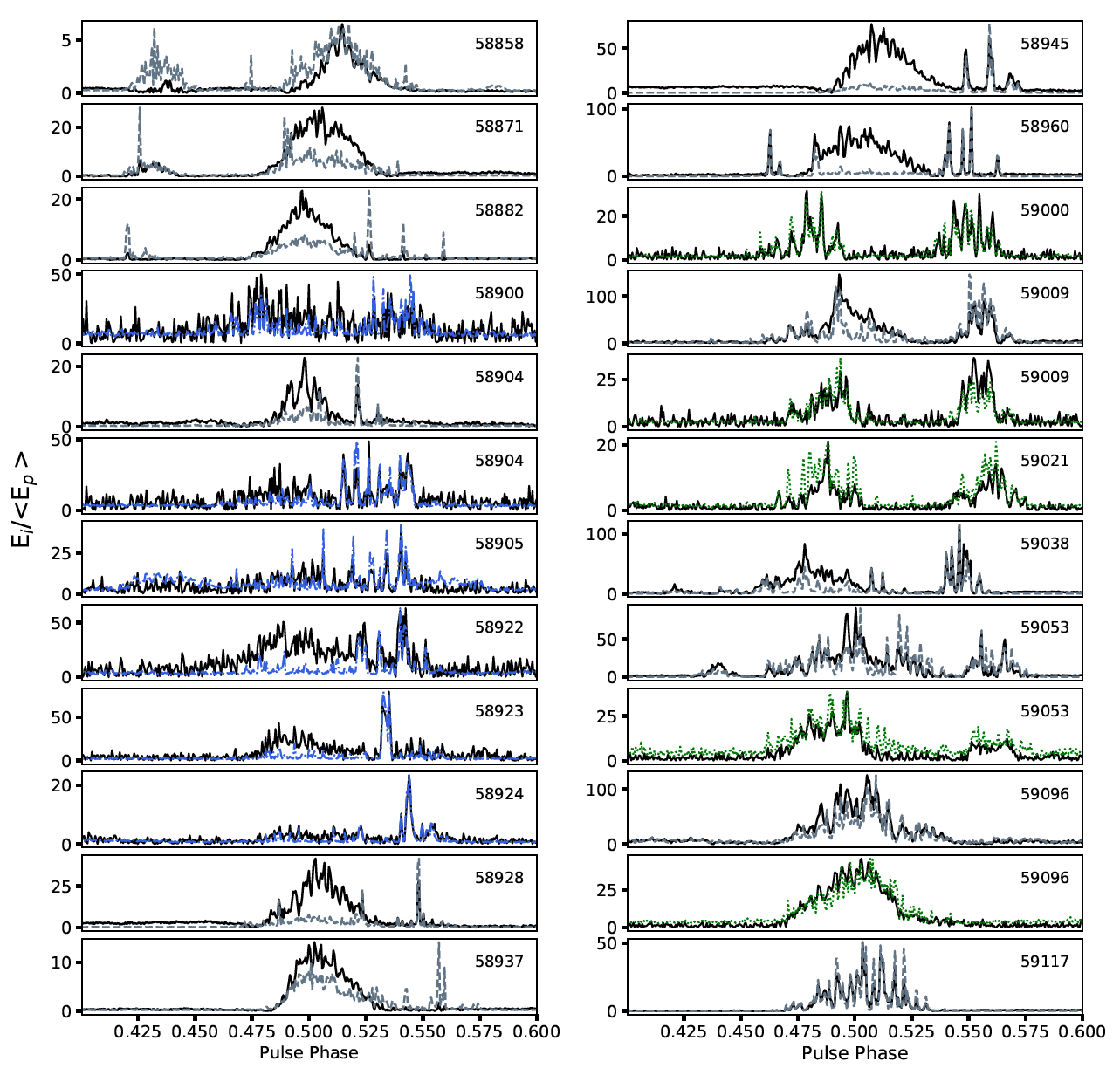}
\caption{Longitude-resolved distributions of the MkII (dotted, green) and Lovell (dash-dot, blue) telescopes at 1564 MHz, and the Effelsberg (dashed, grey) telescope at 6 GHz. The dash-dot, dotted and dashed line shows the ratio of the brightest time sample in each pulse-longitude bin, to the average peak flux $\langle E_{p} \rangle$ at the pulse longitude of the peak of the pulse profile. The solid line is the energy of the average pulse profile scaled to the maximum of the ratio. The profiles have been time-aligned using the best-fit timing solution and templates (see Section \ref{sec:timing} for details). Only the data taken with the Effelsberg and MkII telescopes on MJDs 59009 and 59096 are simultaneous.}
\label{fig:LRCED} 
\end{figure*} 
 

\subsubsection{Giant Pulses}
\label{sec:GPs}

Example GPs from the three telescopes are shown in Figure \ref{fig:GPspectra} and a list of the MJDs corresponding to the GP detections is given in Table \ref{tab:GP_MJDs}. The pulse energy distributions of GPs are typically described by a power-law \citep{LCU+95}, while the more normal emission can be characterized by lognormal statistics \citep[e.g.][]{KSv2010, BJB+12}.
Though the spiky emission reported in \cite{MJS+19} are not GPs by definition, the tails of the distributions of the absolute, and period-averaged peak flux densities are well fit by power-laws. The insufficient number of GPs in our Effelsberg and Lovell data do not allow us to fit their distribution. We fit a broken power-law to the GPs 
in the MkII data in Figure \ref{fig:plfit}. The slope of a power-law fit to the GPs evolves from $-1.6\pm0.1$ at the fainter end with a break at $\sim5.6$ mJy to $-4.8\pm0.3$ at the brighter tail end of the distribution. 

In addition to their energies exceeding $10\langle E \rangle$ and exhibiting power-law energy distributions, GPs are also characterised by their confinement to narrow pulses phase windows. Figure \ref{fig:phasedist} shows the distribution of the phases of all the GPs reported in this work. The bi-modal distribution of the MkII data is a result of the two profile components seen in Figure \ref{fig:LRCED}. We observe 61\% of the MkII GPs to be confined to the leading pulse phase ($\phi$) range $0.473 \leq \phi\leq 0.502$ and the remaining 39\% to be in the trailing phase range $0.541 \leq \phi \leq 0.567$. This preference of the leading over the trailing pulse phase window is similar to what is observed in J0540$-$6919 \citep{GSA+21} and contrary to other systems where the GPs lag the normal emission.

\begin{figure}
\centering
\includegraphics[width=2.8 in]{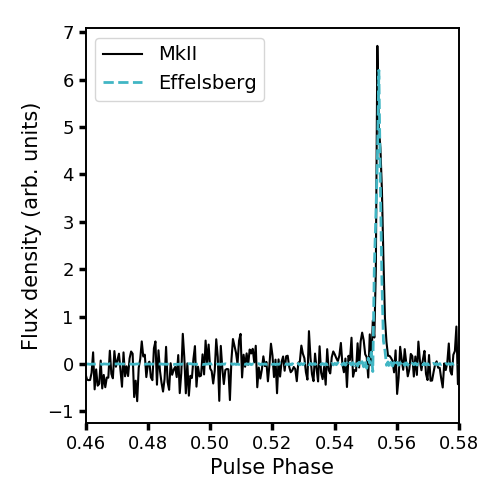}
\caption{Comparison of a GP detected simultaneously with the MkII and Effelsberg radio telescopes. The data are uncalibrated and the flux densities are in arbitrary units. See Section \ref{sec:SpikeyEmission} for details.}
\label{fig:GPcomparison} 
\end{figure}

During the simultaneous observations on 9 June 2020, 13 GPs were detected in the MkII data and 7 GPs were detected in the Effelsberg data. Of the 13 MkII GPs, only 7 occurred during the duration of overlap with Effelsberg observations and all 7 were detected as single pulses in the Effelsberg data. However, only 2 of these 7 simultaneous MKII and Effelsberg pulses were classified as GPs in the both data sets. This is not unexpected given the large stochastic variations possibly due to a varying spectral index \citep{LJK+08, YogeshM, DLB+19, PMP+20} in the underlying normal emission of the integrated profiles used to calculate the average energies. Figure \ref{fig:GPcomparison} shows a comparison of simultaneously detected GPs with the Effelsberg and MkII telescopes. \cite{LJK+08} present simultaneous and quasi-simultaneous multi-frequency observations of XTE~J1810$-$197, and show that though the spectrum is generally flat, large fluctuations of the spectral index is observed with time. Simultaneous dual-frequency observations at 8.3 and 31.9 GHz undertaken in \cite{PMP+20} show spiky emission. Though \cite{PMP+20} do not investigate the existence of GPs, they do note that not all of the emission components were detected at both frequencies suggesting a variable radio spectrum. 

\begin{figure}
\centering
\includegraphics[width=2.5 in]{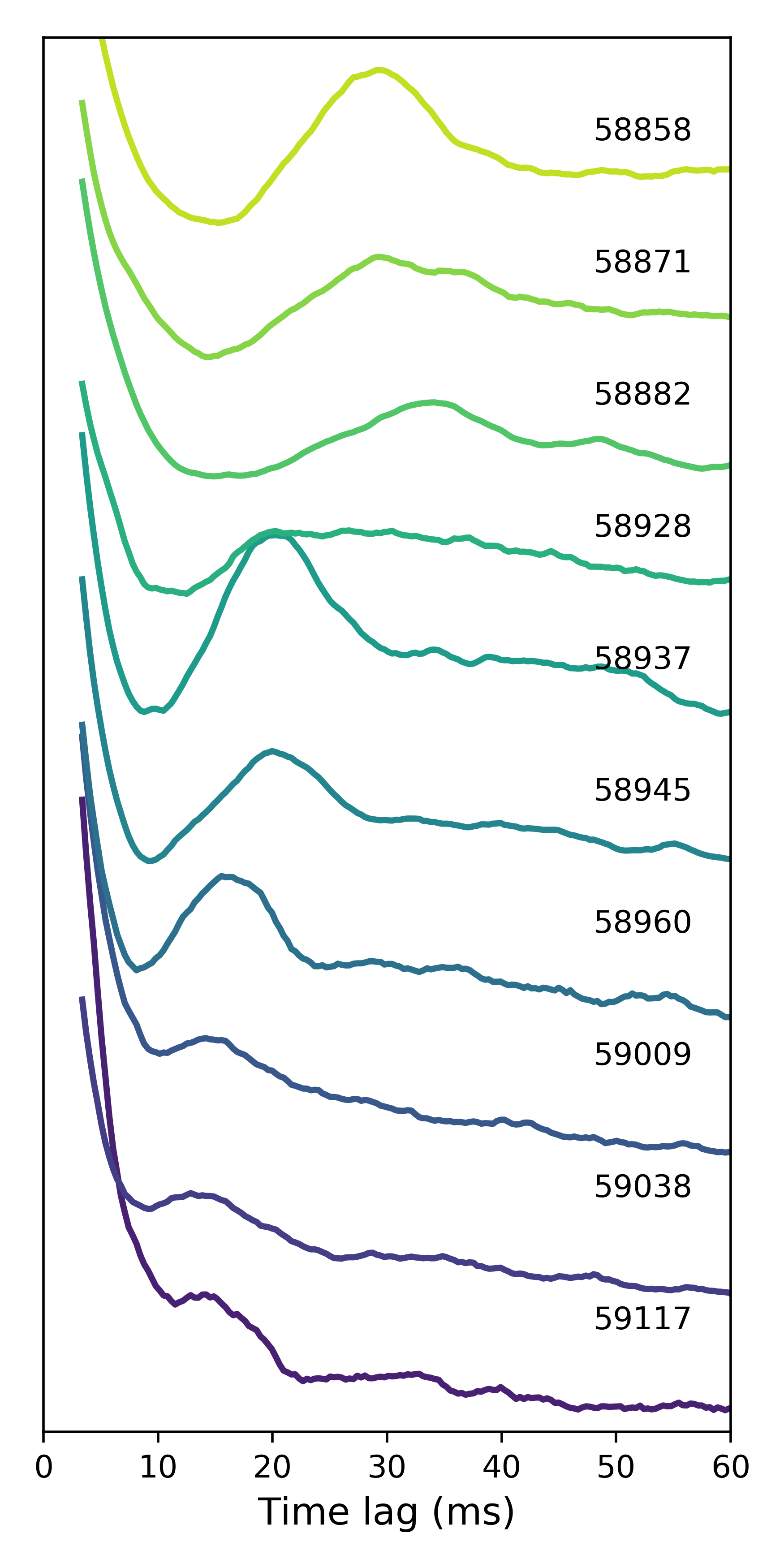}
\caption{Average ACFs of the Effelsberg data. Individual ACFs of the time averaged pulse profiles as a function of time lag are shown for each MJD. The timescale of the characteristic separation between sub-pulses is estimated to be the peak of the first feature following the zero-lag.}
\label{fig:acf} 
\end{figure} 

\begin{figure}
\includegraphics[width=3.5 in]{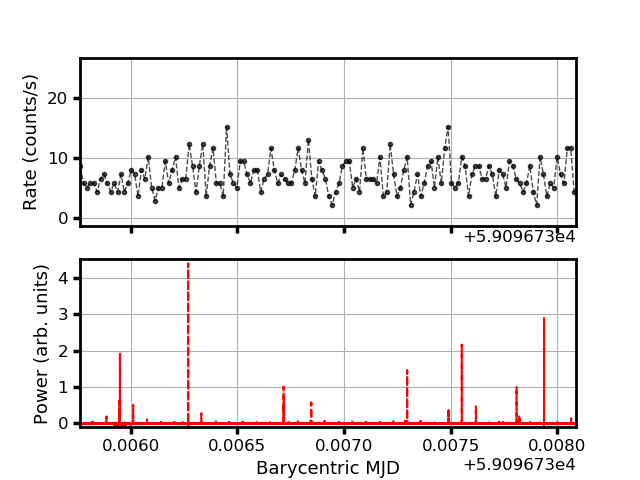}
\includegraphics[width=3.5 in]{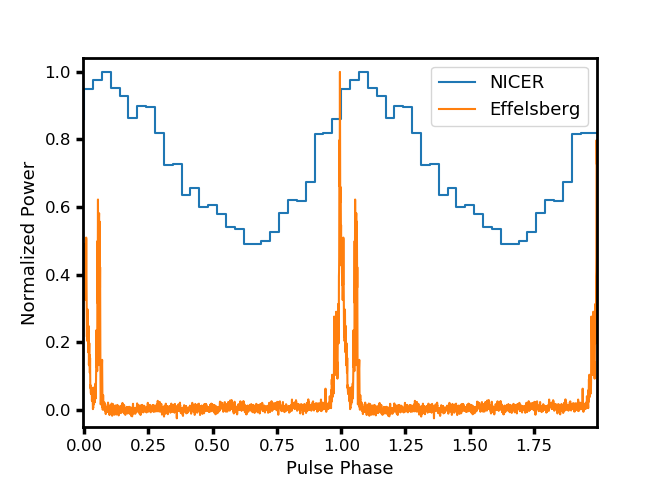}
\caption{\textit{Top Panel:} Simultaneous X-ray (NICER, black dots) and radio (Effelsberg, red) barycentred light curves for XTE~J1810$-$197 for the observation on 09 June 2020. The lack of correlation with the bright radio pulses is evident although the counts per bin in the X-ray light curve are not enough for conclusive evidence (see text for details). \textit{Bottom Panel:} Folded X-ray (blue) and radio frequency (orange) pulse profile for XTE~J1810$-$197 for the same epoch.}
\label{fig:lc} 
\end{figure} 

\begin{figure*}
 \centering
  \includegraphics[width=3.4 in]{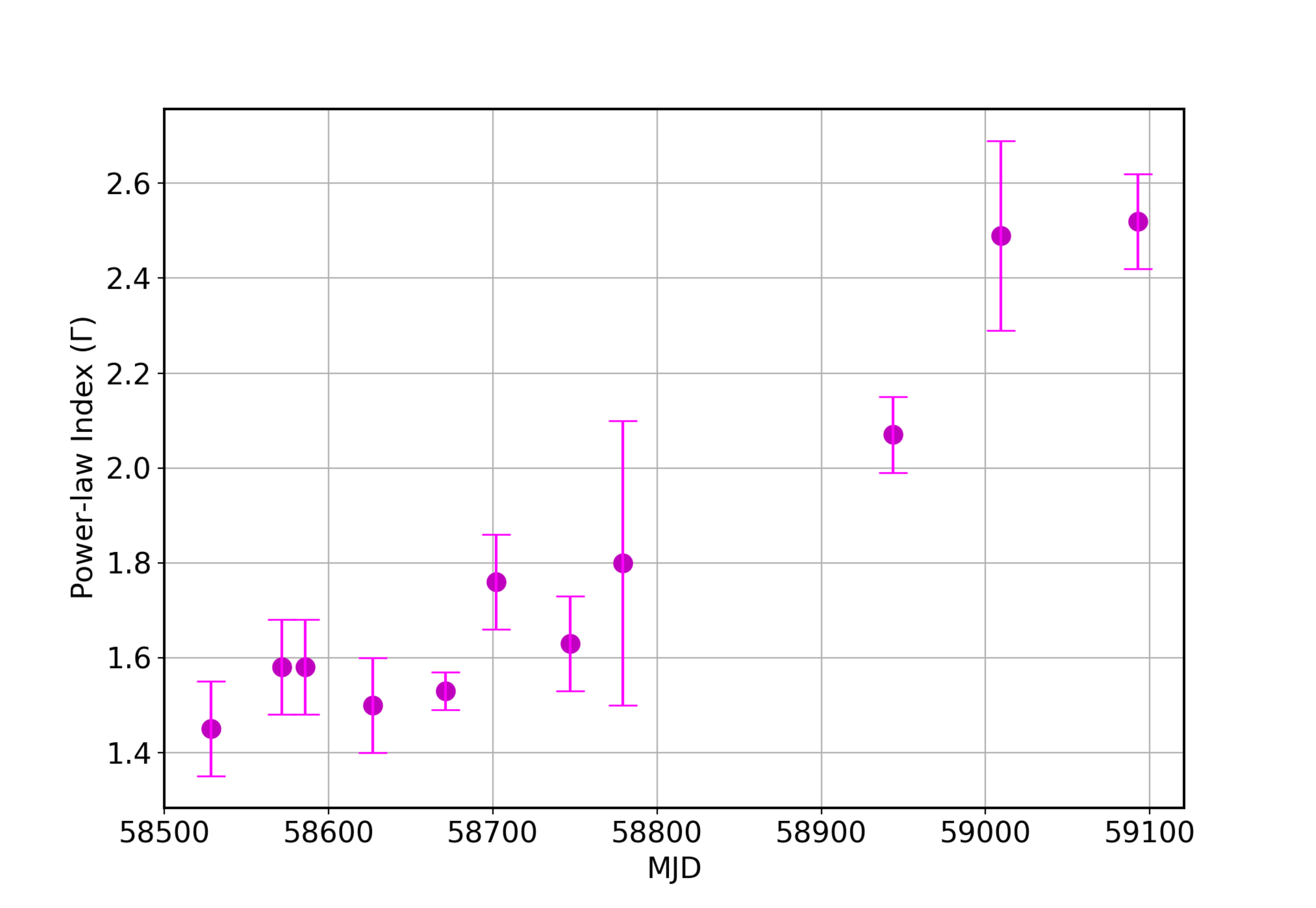}
 \includegraphics[width=3.4 in]{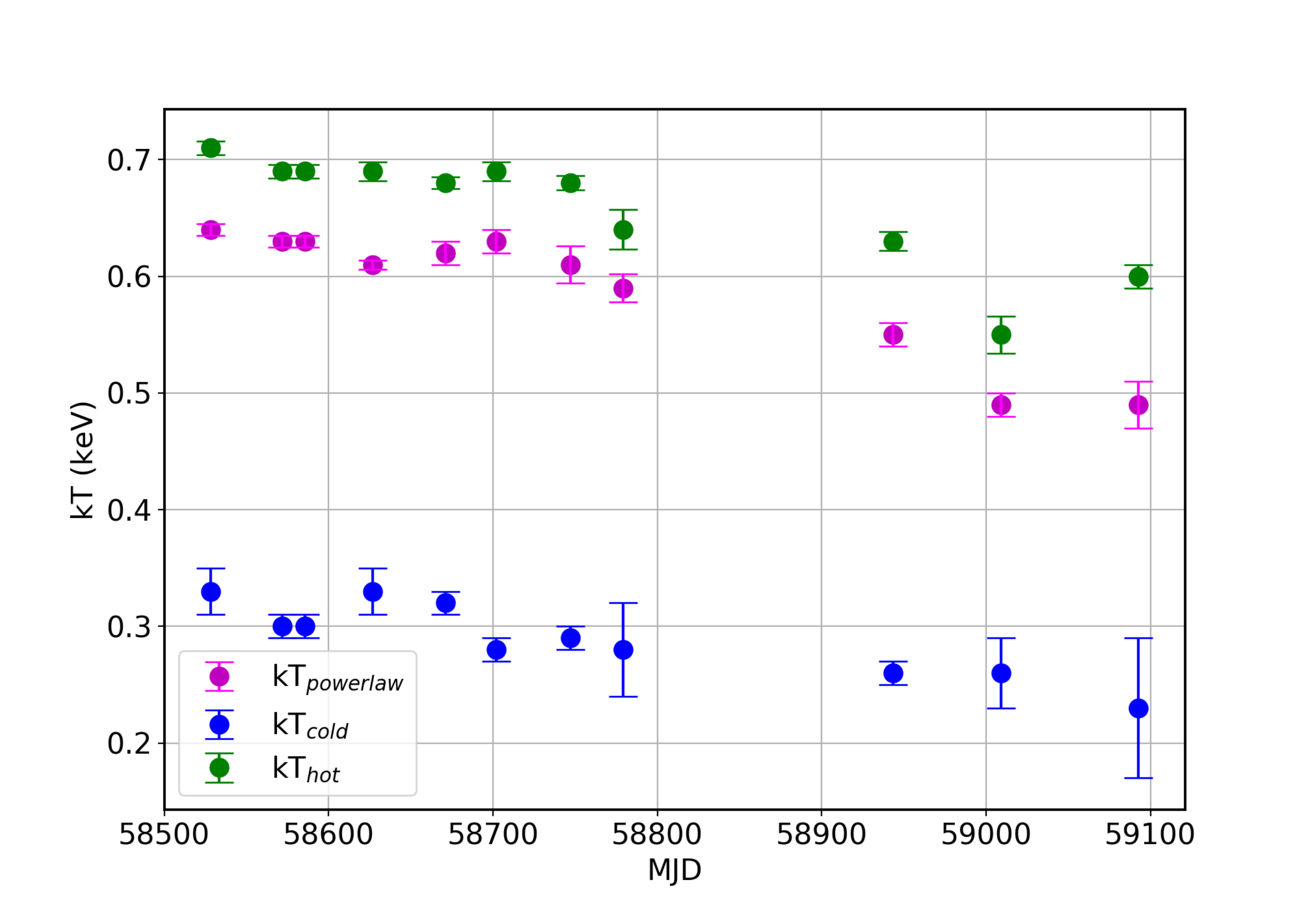}
 \caption{\textit{Left Panel:} Evolution of the power-law index $\Gamma$ as a function of MJD. \textit{Right Panel:} Evolution of the blackbody temperatures as a function of time. $kT_\mathrm{powerlaw}$ is the single blackbody component that was fit along with the powerlaw. $kT_\mathrm{cold}$ is the colder blackbody component and $kT_\mathrm{hot}$ is the blackbody component from the hotspot in the polar-cap where we expect the pulsed X-ray emission to originate from.}
\label{fig:pl} 
\end{figure*} 

\begin{figure}
\includegraphics[width=3.4 in]{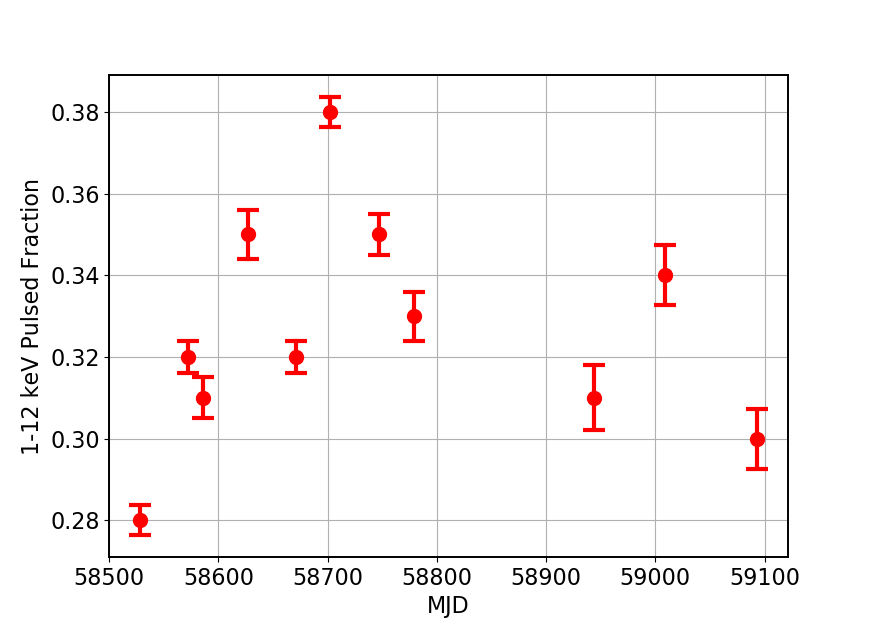}
\caption{The 1--12~keV pulsed fraction from NICER observations of XTE~J1810$-$197 as a function of MJD.}
\label{fig:pf} 
\end{figure}

\subsubsection{Longitude-resolved cumulative energy distributions}
Another phenomenon related to GPs, are giant micropulses \citep{JvK2001} characterised by extremely large peak flux densities but not so much in terms of their integrated pulse energies. They also have a tendency to be confined to a restricted pulse-longitude range of the overall pulse profile. To investigate this, we produced the longitude-resolved cumulative energy distributions shown in Figure \ref{fig:LRCED}. The brightest time sample in each pulse-longitude bin is compared with the peak flux density of the average profile $\langle E_{\rm{p}} \rangle$. We see that the highest measured peak flux of a single pulse is $\sim 150\langle E_{\rm{p}} \rangle$ at MJD 59009 in the Effelsberg data. 
We see that GP-like pulses first appeared on MJD 58871, confined to the leading edges of the precursor to the left of the main component of the pulse profile. They then however, switched positions midway through the observation on 58882 from the left of the main component to its right, and remained on the right of the main pulse until 58945. On 58960 the pulse profile was much spikier than before with GP-like emission at confined phase ranges on either side of the main component, in addition to the leading edge of the main component. Between 59000 and 59117 inclusive, the GP-like emission made an appearance across the entire phase range occupied by the main component, in addition to the component to the right of the main component. As of 59117, the average pulse profile emission was at its spikiest yet, with the fluctuations in the main pulse much more distinct than the other days.

\begin{figure}
\includegraphics[width=3.3 in]{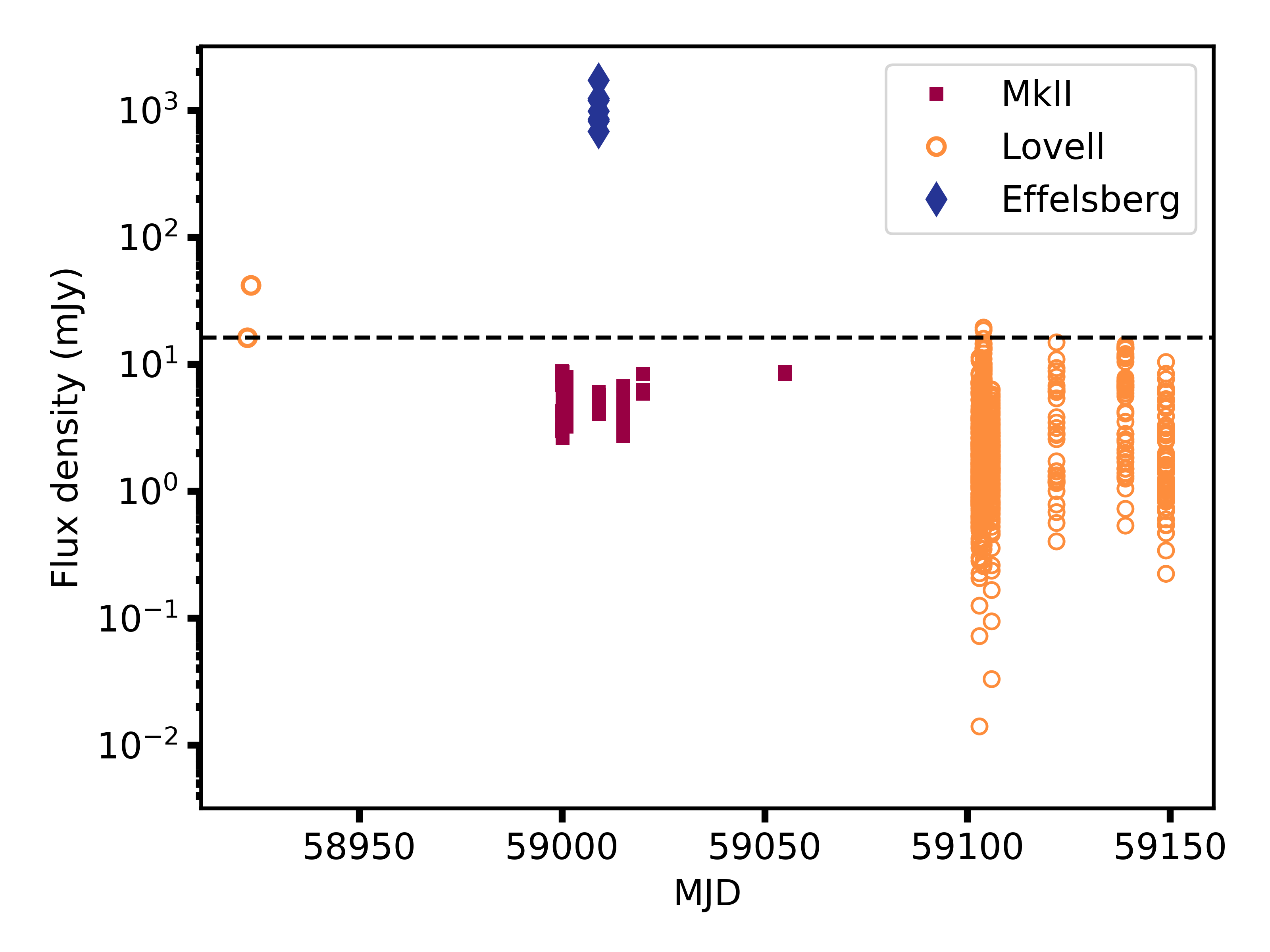}
\caption{Average fluxes of the GPs detected by the Effelsberg, Lovell and MkII telescopes as a function of MJD. The dashed line marks the flux of a GP at the Lovell telescope whose E/$\langle E \rangle$ $\sim 12$. At least two of the pulses from the Lovell data between MJDs 59100 and 59150 represented by open circles, are potentially GPs even though they do not meet the criterion in Figure \ref{fig:pulsedist}. This is due to the increase in flux of the underlying emission in the recent Lovell data.}
\label{fig:GPfluxes} 
\end{figure} 

\begin{figure}
\includegraphics[width=3.5 in]{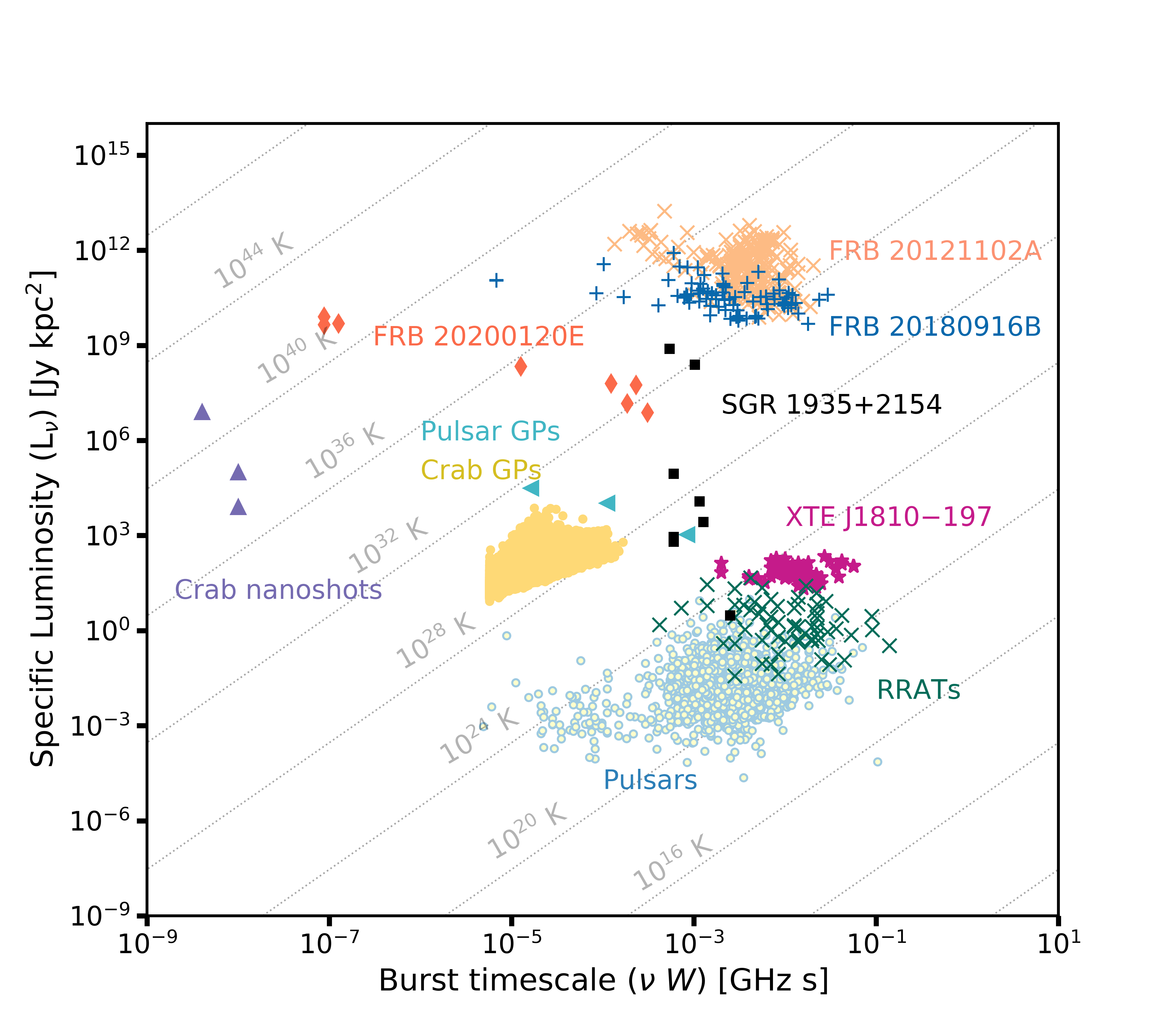}
\caption{Specific luminosity as a function of burst timescale of coherent radio sources which emit nanosecond to millisecond duration pulses. The FRBs plotted here are repeating FRBs with well known distances, and SGR~1935$+$2154 is the radio-loud Galactic magnetar which produced an FRB-like pulse. GPs from XTE J1810$-$197 are seen to sit above the clusters of pulses from rotating radio transients and canonical radio pulsars, forming their own distinct cloud. Figure adapted from \citet{NHK+21}. }
\label{fig:phasespace} 
\end{figure}

\begin{figure*}
\includegraphics[scale=0.8]{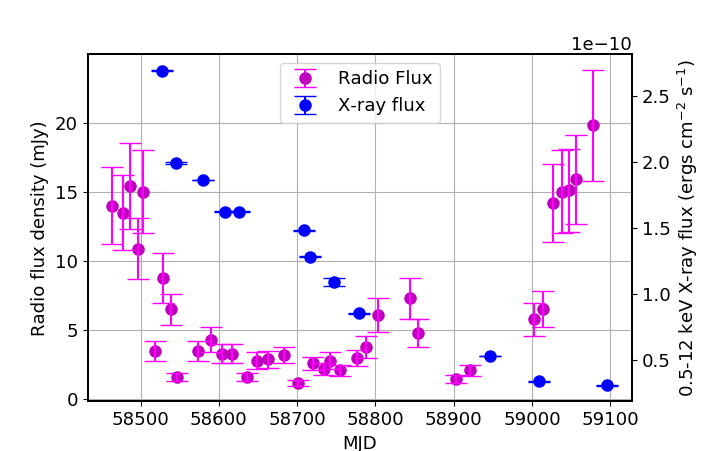}
\caption{Evolution of radio (1564~MHz) and 0.5--12~keV X-ray flux as a function of time for XTE~J1810$-$197 since the onset of the most recent outburst. Note that the X-ray flux has been computed assuming the two blackbody model for the X-ray spectrum (see Table~\ref{tab:specparams}).}
\label{fig:flux_evo} 
\end{figure*}

\subsection{ACF Analysis}

Substructure in single pulses from pulsars often manifests with two distinct timescales, and potentially origins. There are the sub-pulses which tend to have widths which are between half and a tenth of the average pulse width or $\sim$1-20 ms. Then there is microstructure which are characterised by timescales of $\sim10\,\upmu$s - 4 ms \citep[e.g.][]{Lange1998}, which can appear superimposed on top of the sub-pulses, or make up the sub-pulses. When one takes an autocorrelation function (ACF) of pulses containing both sub-pulses and microstructure, one will see a peak at zero-lag corresponding to the DC component, then a peak at short timescales due to the microstructure and then a second peak associated with the sub-pulses \citep{CWH1990}.  Visual inspection of the single pulse profiles of the Effelsberg data exhibit spikes which are not only narrow, but appear to be quasi-periodic in nature. We performed an ACF analysis of the single pulse intensities \citep{Lange1998} using the data with $\sim 340\,\upmu\rm{s}$ time resolution to test this hypothesis. We compute the cross-correlation of the de-dispersed signal as a function of time with a delayed copy of itself given by,

\begin{equation}
    \mathrm{ACF}(\tau) = \int_{0}^{t} f(t) f(t-\tau) \, \mathrm{dt} ,
\end{equation}

\noindent where $\tau$ is the time lag. The zero-lag value, associated with self-noise, was excised from the ACF. In general, the peaks in the zero-lag excised ACFs are subtle except in some cases where prominent peaks are observed. Consequently, the single pulse ACFs for each MJD were averaged. Figure \ref{fig:acf} shows the normalised averages for each MJD. 
The characteristic separation of peaks in the structure of single pulses is given by the time lag of the peak of the first feature at non-zero lag shown in Figure \ref{fig:acf}. On average we observe the separation between sub-pulses to decrease as a function of MJD.

\begin{table*}
\caption{The 0.5--12~keV spectral parameters for XTE~J1810$-$197 for a 2BB model and the PL model. The X-ray flux corresponds to the unabsorbed flux.}
\label{tab:specparams}
\centering
\begin{tabular}{l l l l l l l l}
\hline \\ 
Date & $kT_{\rm warm}$ & $kT_{\rm cold}$ & $\Gamma$ & $F_{\rm X, PL}$ (0.5--12~keV) & $F_{\rm X, BB}$ (0.5--12~keV)  & Pulsed Fraction & Exposure time\\ [1ex]
(MJD) & (keV) & (keV) & & ($\times$10$^{-10}$ ergs~s$^{-1}$) & ($\times$10$^{-10}$ ergs~s$^{-1}$) & ($\%$) & (s)\\ [1ex]
\hline \\
58522.02078704 & 0.75$\pm$0.01 & 0.38$\pm$0.01 & 1.52 $\pm$0.04 & 2.450$\pm$0.005  & 2.690$\pm$0.003 & 28 $\pm$1.1  & 4826.0 \\
58571.80939815 & 0.70$\pm$0.01 & 0.32$\pm$0.01 & 1.74 $\pm$0.05 &1.770$\pm$0.004 & 1.990$\pm$0.005& 32$\pm$1.2 & 4552.0\\
58585.67546296 & 0.702$\pm$0.01 & 0.32$\pm$0.01 & 1.76$\pm$0.05 &1.690$\pm$0.003 &1.860$\pm$0.001 & 31$\pm$1.5 & 4930.0 \\
58626.90393519 & 0.71$\pm$0.01 & 0.33$\pm$0.01 & 1.80$\pm$0.05 &1.470$\pm$0.003 &1.620$\pm$0.004 & 35$\pm$1.8 & 3529.0 \\
58702.02623843 & 0.74$\pm$0.01 & 0.34$\pm$0.01 & 1.77$\pm$0.05 &1.310$\pm$0.002 &1.620$\pm$0.004 & 38$\pm$1.1 & 3934.0\\
58747.18009259 & 0.69$\pm$0.01 & 0.30$\pm$0.01 & 1.88$\pm$0.04 &1.09$\pm$0.03 &0.954$\pm$0.003 & 35$\pm$1.5 & 7004.0\\
58779.11630787 & 0.67$\pm$0.02 & 0.29$\pm$0.03 & 2.01$\pm$0.20 &0.851$\pm$0.005 &0.790$\pm$0.003 & 33$\pm$1.8 & 9290.0 \\
58943.58254630 & 0.63$\pm$0.01 & 0.26$\pm$0.01 & 2.33$\pm$0.08  &0.498$\pm$0.002 &0.527$\pm$0.001 & 31$\pm$2.4 & 4541.0 \\
59009.14583333 & 0.58$\pm$0.01 & 0.27$\pm$0.02 & 2.33$\pm$0.20 &0.302$\pm$0.002 &0.336$\pm$0.002 & 34$\pm$2.2 & 4682.0 \\
59092.78406250 & 0.64$\pm$0.01 & 0.24$\pm$0.01 & 2.36$\pm$0.07 &0.262$\pm$0.001 & 0.304$\pm$0.001 & 30$\pm$2.2 & 1684.0 \\
\hline
\end{tabular}
\end{table*}

\section{X-ray Data Analysis and Results}

\subsection{X-ray Lightcurves}

The radio and the X-ray lightcurves were time-stamped with reference to the Solar System barycentre and the resulting light curve showing overlapping time intervals (see section 3) is shown in the top panel of Figure~\ref{fig:lc}. It is visually evident from the plot that there is no obvious correlation between the X-ray and the radio emission. Identifying any underlying correlation between the two wavelengths is more difficult than the previous simultaneous radio and X-ray observations presented in \cite{PMP+20} as the X-ray photon counts are at least a factor of 4 lower compared to their data which were closer to the onset of the outburst. Both lightcurves were folded modulo the topocentric period derived from the best fit ephemeris shown in Table~\ref{tab:timingparams}. The resulting folded X-ray and radio profiles are shown in the bottom panel of Figure~\ref{fig:lc}. The X-ray profile is best characterized by a simple sine wave with no hint of other components which is dissimilar to the radio profile that shows multiple components within the emission region. It is also important to note that the peak of the X-ray profile is offset from the peak of the leading component of the radio profile by approximately 0.1 in pulse phase which is consistent with the offset seen at the onset of the outburst in 2018~\citep{Gotthelf2019}. Such an offset between the X-ray and radio pulse is commonly seen in other radio loud magnetars and is expected if the X-ray emission is coming from a different part of the magnetosphere than the open field lines where the radio emission originates from~\citep{Gotthelf2019}. 

\subsection{X-ray Spectroscopy}

In order to study the long-term spectral evolution of XTE~J1810$-$197, each observation shown in Table~\ref{tab:specparams} was analyzed separately. As the background dominates the spectrum below 0.6~keV and the X-ray flux from the source is extremely weak above 4.5~keV, we only fit the spectra in the 0.6--4.5~keV range. For the spectral fitting, we used the \textsc{tbabs} model in \textsc{xspec} \citep{Arnaud} assuming interstellar medium (ISM) abundances \citep{Wilms}. The spectrum was binned so that each spectral channel had at least 25 counts to improve the S/N per spectral channel and facilitate a better fit to the data. Previous spectral studies of XTE~J1810$-$197 have shown that a two-component blackbody model and a combination of a powerlaw and a single blackbody component can explain the soft X-ray spectrum~\citep{GH2005}. However, the power-law is the dominant component in the softer part of the spectrum and there is no physical model that can properly explain or reconcile  with the spectrum seen at optical/infra-red wavelengths~\citep{gotthelf2007}. Regardless, we decided to fit both (powerlaw + blackbody and blackbody +blackbody) models separately to the spectra to look for the evolution of parameters as a function of time. To search for any statistically significant variations in any of the fitted parameters, we fit for all parameters by keeping only the neutral Hydrogen column density (N$_\mathrm{H}$) as a constant. For the purpose of this analysis, we used a N$_\mathrm{H}=6.5\times$10$^{21}$~cm$^{-2}$ \citep{GH2005}. For each observing segment, we fit the spectrum with both the aforementioned models and the resulting best fit parameters are shown in Table~\ref{tab:specparams}. We then analyzed the evolution of the best-fit parameters since the onset of the outburst. We clearly see an evolution of our fitted parameters with time and we discuss the implications of this variation in Section \ref{sec:R&D} (see Figure~\ref{fig:pl}). We should note that~\cite{borghese2021} used a 4 blackbody component model to fit the NICER data but the coolest and the hottest components were fixed using the best fit values from the XMM-Newton and \textit{Nu-STAR} data respectively. Since we only use the \textsc{NICER} data, we decided to fit it with a two blackbody component model. We note that while we have used a similar dataset as~\cite{borghese2021}, we have analysed the data for a much larger baseline compared to what is presented in~\cite{borghese2021} with an extension of almost another year. We discuss the implications of the results in the next sections.

\subsection{X-ray Pulsed Fraction}

 During the previous outburst of XTE~J1810-197, the 1--15~keV pulsed fraction decreased from 55$\%$ at the onset of the outburst to 25$\%$ two years after the outburst. After this it remained approximately constant until the recent reactivation of the magnetar~\citep{halpern2005, bernardini2009}. The high cadence NICER observations give us an opportunity to study the evolution of the pulsed fraction since the onset of the most recent outburst. Because of the high particle background in some of the NICER observations in 0--1~keV range, we decided to compute the pulsed fraction in the 1--12~keV range to avoid contamination of the spectrum by spurious photons from the Sun. For each folded X-ray profile, we calculated the pulsed fraction
\begin{equation}
\eta = \frac{\sum_{\rm n=1}^{\rm nbins} \left( F_{\rm n} - \rm min(F)\right)}{\sum_{\rm n}^{\rm nbins} F_{\rm n}},
\end{equation}
where $F_{\rm n}$ is the flux in a given phase bin~\citep[see][]{chen2008}. Figure~\ref{fig:pf} shows the evolution of the 1--12~keV pulsed fraction over time and a very different picture to the previous outburst is seen. The pulsed fraction in the 1--12~keV range increases from 28$\%$ to 38$\%$ in the immediate aftermath of the outburst while showing significant scatter in that trend. The pulsed fraction peaks at MJD~58700 before gradually decreasing and seems to come back to its original value of 28$\%$ at the start of the outburst.

\section{Discussion}
\label{sec:R&D}

XTE~J1810$-$197 has been regularly monitored and timed with the Lovell, MkII and Effelsberg telescopes from 2009 \citep{LLD+19} to date, to study the long term evolution of the spin-down rate of the magnetar. Since the middle of 2019 to the beginning of 2020, there has been a steady decrease in the spin-down rate $|\rm{F1}|$ shown in Figure \ref{fig:timing}, reaching a minimum in the beginning of 2020, followed by an increase resulting in a doubling of $|\rm{F1}|$ since then. Interestingly, this minimum in $|\rm{F1}|$ corresponds to the onset of the spiky GP-like emission discussed in Section \ref{sec:SpikeyEmission}. The overall amplitude and structure in the evolution of $|\rm{F1}|$ since the second outburst in 2018 is initially very similar to the one reported in \cite{CRH+16} and \cite{LLD+19} following the first outburst in 2003. This part of the recovery is also reminiscent of the what has been observed in the spin-frequency recovery of high magnetic field pulsars after an X-ray outburst \citep{DJW+18}. This suggests that high magnetic field pulsars and magnetars plausibly share similar spin-frequency evolutionary traits. 

From Figure \ref{fig:timing}, we can see that F1 changes significantly over the course of the recovery from the outburst. This is fairly typical of radio-loud magnetars \citep{KB+17}, but unlike what is seen in canonical radio pulsars where F1 remains fairly constant or shown variations of much smaller magnitude \cite[see][and references therein]{LBS+20}. Hence, one needs to be careful while interpreting inferred quantities like the surface dipole magnetic field strength (B) and the characteristic age ($\tau$) of the magnetar from the measured F0 and F1. The inferred B changes by a factor of 3 while $\tau$ changes by a factor of 2 over the time span of our observation thus, demonstrating the issue.

It is unknown at what time after the 2003 outburst the radio pulsations turned on. We note that the time covered by the analysis presented here for data obtained since the 2018 outburst, is less than the number of days between the 2003 X-ray outburst and the radio emission turn-off. It is also less than the discovery of pulsed radio emission in the first outburst and the radio emission turn-off. However, we have continued to monitor the source till date, and as of August 2021, it is still visible as a radio emitter after $\sim$1000 days. 
It will be interesting to see if the trend in the spin-down evolution from the first outburst is repeated, with the current erratic spin-down behaviour reaching a stable minimum $|\rm{F1}|$ followed by radio pulsation turn off within the next $\sim$1000 days.

The pulse energy distributions of the data from the three telescopes at 1564 MHz and 6 GHz show radio pulses with energies greater than 10$\langle E \rangle$, which qualifies them as GPs. This definition of a GP is heavily dependent on the average pulse energy of the underlying emission. In ordinary pulsars this average pulse energy remains fairly constant over several days. However, we observe a systematic variation in this underlying emission in magnetars, thereby affecting the interpretation of these bright pulses as GPs. 
While the average pulse energy estimation is based on observations covering MJDs for the Effelsberg data, the average pulse energy per observation was used for the Lovell and MkII data due to the large systematic variations in the underlying emission over a timescale of a few days. We note that for the more recent Lovell data from MJDs 59102, 59103, 59104, 59106, 59118, 59122, 59139 and 59149, we see the appearance of bright normal emission in almost all single pulses in addition to the GP-like emission. While we do not see $E/\langle E \rangle$ exceed 10 for any of these pulses, we identify single-peaked pulses from MJD 59104 that have similar flux densities to the GPs seen in earlier Lovell data (see Figure \ref{fig:GPfluxes}). These pulses are not GPs when defined by comparing their flux to that of the increasing flux density of the underlying emission. However, they are clearly drawn from the same pulse class or type.

The spectra of some magnetars including XTE~J1810$-$197 \citep[e.g.][]{LJK2+2008}, are known to vary between somewhat positive and negative spectral indices \citep[e.g.][]{TDE+17}.
This is reminiscent of FRB pulses \citep[e.g.][]{nat_spitler, Pastor-Marazuela} as well as Crab GPs \citep{KSv2010}, which show wildly varying spectra with some pulses brighter at higher frequencies. However, unlike what has been observed in most repeating FRBs, we do not see any evidence of downward frequency drifting with progressing time \citep{HSS+19, CSA+20}. While the number of $E >10\langle E \rangle$ GPs detected in the Effelsberg and Lovell data are too low to fit their distribution, a broken power-law was found to describe the MkII GPs with a slope of $-1.6 \pm 0.1$ at the fainter end and $-4.8 \pm 0.3$ at the bright end of the distribution. The power-law index of the faint end of the distribution is comparable to those measured for the repeating FRB~20121102A by \cite{GMS+19} ($-1.8\pm0.3$) and \cite{OMvL+20} ($-1.7\pm0.6$). The power-law index of the bright end of the distribution lies between that of GPs emitted by the Crab at 1.3 GHz $(\alpha = -3.36)$ \citep{Lundgren} and giant micro-pulses emitted by PSR B1706$-$44 at 1.5 GHz $(\alpha = -6.4\pm0.6)$ \citep{Cairns2004}.

As there is growing evidence for the association of at least some FRBs being from (young) Magnetars \cite[e.g.][]{MM18}, especially because of the exceptionally bright radio pulse seen from the Galactic magnetar SGR~1935$+$2154 \cite{CHIME_SGR, brb+20} it is worth comparing the brightest of the GPs at all frequencies from XTE~J1810$-$197 with those from SGR~1935$+$2154 and the FRBs. In Figure \ref{fig:phasespace} we compare the specific luminosity of a sample of the GPs seen from XTE~J1810$-$197 with a whole range of coherent pulsed radio emitters. We can see that the GPs lie well above the normal pulsars and the majority of the rotating radio transients (RRATs) and are comparable to some of the Crab GPs. They are however still 1-2 orders of magnitude fainter than some of the pulses from the Galactic magnetar SGR~1935$+$2154. We note that the pulse widths used in the estimation of the luminosities of the XTE~J1810$-$197 GPs are the observed and not intrinsic widths. As the time resolution is constrained by the sampling time and the scatter broadening seen in both the 6~GHz and 1564~MHz pulses, it means that the GPs could be somewhat narrower and thus move to the left and up in this diagram.

The timescale of the quasi-periodic modulation is observed to be variable over a timescale of days (Figure \ref{fig:acf}).
In the Effelsberg observations on most MJDs, micropulses are visually evident in the single pulse profiles in the form of oscillations superimposed on an envelope, but either do not always feature in the corresponding ACFs or translate into very subtle oscillations. This could be due to either a small number of micropulses in the single pulse ACFs causing them to be washed out in the average ACF, or the absence of any preferred micropulse duration as shown in \cite{CWH1990}. The decrease in the separation between sub-pulses as a function of MJD in the Effelsberg data is consistent with the evolution of the pulse profile in Figure \ref{fig:LRCED}. 

During the radio monitoring campaign, the magnetar was also regularly monitored by NICER. This gave us an opportunity to study the spectral evolution of the source since the outburst and if this evolution is in any way related to the radio properties of the magnetar. During the previous outburst in 2003, the X-ray spectrum of the source was well described by a three temperature black-body model with the hottest component proposed to be coming from a small hot spot on the surface and a warmer component coming from another hot spot over a larger surface area and the coldest component coming from the cooling of the entire neutron star surface~\citep{bernardini2009}. It is also possible to explain the spectral evolution of XTE~J1810$-$197 with a two component blackbody model without a need of the cool component from the entire surface of the neutron star~\citep{gotthelf2007}. Our two component blackbody model fits to the NICER data show that the evolution of the two hot spots on the surface of the neutron star follow a similar trend compared to the previous outburst (Figure~\ref{fig:pl}). The temperature remains approximately constant for a year before starting to cool down gradually. We also note that the values of the warm and hot component are consistent with the analysis presented in~\cite{borghese2021} with the same dataset and are also consistent with those estimated by~\cite{gotthelf2007} from XMM-Newton observations of the previous outburst. We also show the evolution of the two blackbody components during the time when GPs were observed. We see the same trend of continued cooling of the hot spot during this time. This timescale of decay can be explained by the coronal heating model~\citep{beloborodov2006} in which a starquake causes charged particles to be accelerated through a twisted magnetic flux tube while collision of charged particles with the neutron star surface creates the hot spots. While the overall X-ray spectrum is consistent with the previous outburst, the main difference is the complete lack of the 1.1~keV absorption feature that was reported by~\cite{bernardini2009}. Typically, absorption features in the X-ray spectra of magnetars can be attributed to proton cyclotron absorption in the magnetosphere that is very much dependent on the magnetic field and the plasma density~\citep{TE+13}. To make sure that the absence of this absorption feature was not due a lack of sensitivity, we looked at NICER data close to the outburst where the 0.6-10~keV X-ray flux was a factor of 10 higher than the 0.6-10~keV flux of 5$\times$10$^{-12}$~ergs~cm$^{2}$~s$^{-1}$ when the 1.1~keV feature was detected during the previous outburst. We do not detect such a feature in any of the NICER datasets. This can be attributed to a couple of reasons. One could be that the XMM-Newton observations were more sensitive to this feature. On the other hand, the feature could be absent due to the different geometry observed during the current outburst. The change in geometry is a likely scenario as many authors have suggested a shift in the magnetosphere of the neutron star due to a starquake if we assume that starquakes cause outbursts in magnetars~\citep{belobodorov2007}. While the phase lag between the radio and the X-ray profiles has not changed compared to the previous outburst~\citep{Gotthelf, borghese2021}, the evolution of flux at radio wavelengths differs significantly compared to the previous outburst, suggesting an evolution in the location of the X-ray/radio emitting regions caused due to a change in the region traversed by the line of sight. This argument is also supported by the evolution of the pulsed fraction observed during the current outburst, which is significantly different from what was observed in the previous outburst (Figure~\ref{fig:pf}). The value and the evolution of the pulsed fraction depends on the interplay between the line-of-sight and the location of the hot spot and how the area of the hot spot changes with time. We acknowledge that we cannot rule out the possibility that the different evolution we see in this outburst is caused by the emitting region originating from a different part of the polar cap. Studies of the evolution of the polarisation of radio emission from the magnetar will provide a more direct evidence for one of the two possibilities (Desvignes et al., in prep).

The most recent outburst of XTE~J1810$-$197 has been extensively monitored at radio and X-ray wavelengths. While most of these observations are not simultaneous, the span of the monitoring campaigns gives us an opportunity to study the long term evolution of the X-ray and radio flux of the magnetar since the onset of the outburst. Figure~\ref{fig:flux_evo} shows that the evolution is quite different at the two wavelengths. The X-ray flux declines monotonically as expected from the cooling blackbody temperatures that were estimated from the spectral analysis. While the radio flux also initially declines after the outburst, 
around MJD~58800, the radio flux starts to increase again. The radio flux decreased again before the sudden up turn within a span of three months only to rise again back to the fluxes that were seen at the beginning of this monitoring campaign close to the X-ray outburst. During all this time, the X-ray flux kept decreasing showing no correlation whatsoever with the radio flux. The radio flux density in Figure~\ref{fig:flux_evo} seems to roughly track the spin-down rate F1 (or effective magnetic field), which is arguably also large near the start and end of our observations with a minimum near the middle. Such large variations in the radio flux have not been observed in the previous outburst when the radio flux kept decreasing~\citep{CRH+06}. Although, it did show erratic changes in the radio flux for a few weeks before abruptly disappearing as a radio pulsar altogether in 2008 \citep{CRH+16}.


\section{Conclusions}
\label{sec:conclusions}

We report on the detections of GPs, long-term timing properties, and X-ray and radio fluxes during quasi-simultaneous radio and X-ray monitoring of the magnetar XTE~J1810$-$197. The GPs made their first appearance on MJD 58922 during observations with the Lovell radio telescope at 1564 GHz and were continued to be seen through to MJD 59054 during observations with the Effelsberg radio telescope at 6 GHz. We report a total of 91 GPs, which satisfy the classic criterion of their flux density exceeding the average flux density by a factor 10 
with the Lovell, MKII and Effelsberg radio telescopes. The onset of the GPs corresponds to a minimum in |F1|. The pulse phases of these GP are confined to narrow ranges, reminiscent of the GP phase windows observed in other sources of GPs. These GPs are comparable in luminosity to the giant radio pulses seen in pulsars, and are 1-2 orders of magnitude weaker than the lowest energy pulses detected from the Galactic magnetar SGR~1935$+$2154. We continue to see these bright and narrow pulses but by a definition which compares them to the average energy they are no longer strictly GPs. However we note that this definition fails when all the emission is actually due to GPs, which could be the case sometimes in this source. The systematic variation in the underlying emission of XTE~J1810$-$197 between MJDs 59102 and 59149 makes the brightest pulses in that time interval weak compared to the the classic criterion of a GP, while they are clearly drawn from the same population of GPs detected at other times.
The long term X-ray evolution of XTE~J1810-197 shows that the two blackbody components that best describe the NICER spectrum cool down over time. This is consistent with the previous X-ray outburst shown by this source. At the same time, the 0.5--12~keV X-ray pulsed fraction shows a very different evolution compared to what was seen during the first outburst. While the X-ray flux kept decreasing since the outburst, the radio flux tends to show a slight peak around MJD~58850 followed by a rapid increase in flux that aligns with the onset of GPs. We encourage continued radio and X-ray monitoring of XTE~J1810$-$197 in the event of it producing an FRB-like pulse, similar to SGR~1945$+$2154. 

\section*{Acknowledgements}
MC, KR, BWS and MPS acknowledge funding from the European Research Council (ERC) under the European Union's Horizon 2020 research and innovation programme (grant agreement No 694745). The authors would like to thank the staff of the Lovell telescope, Effelsberg telescope and the NICER mission for help with the observations. Pulsar research at the Jodrell Bank Centre for Astrophysics and Jodrell Bank Observatory is supported by a consolidated grant from the UK Science and Technology Facilities Council (STFC).

\section*{Data availability}

The data underlying this article will be shared on reasonable request to the corresponding author.




\bibliographystyle{mnras}
\bibliography{refs} 



\appendix

\section{List of GP detections with the Effelsberg, Lovell and MkII telescopes.}



\begin{table}
\caption{GP detections at the Lovell, Effelsberg and MkII telescopes. The MJDs reported here are within a period of the magnetar.}
\label{tab:GP_MJDs}
\centering
\begin{tabular}{l c c c}
\hline\hline
Telescope  &  Topocentric MJDs & S/N & Width (ms) \\ [0.5ex] 
\hline
Lovell  & 58922.3907102877  & 361.85 & 8.84\\
        & 58923.3009874952 & 580.63 & 11.88\\
\hline
Effelsberg  & 59009.0381518352 & 2965.79 & 9.47 \\  
            & 59009.0387291039 & 1892.83 & 6.08 \\
            & 59009.0403326278 & 2961.27 & 7.05 \\
            & 59009.0411664603 & 4142.83 & 4.54\\
            & 59009.0448866361 & 2855.32 & 9.51\\
            & 59009.0532249615 & 3094.75 & 5.24\\
            & 59009.0614350054 & 3704.77 & 7.03\\
\hline
MkII    & 59000.0599005787 & 58.71 & 13.77 \\
        & 59000.0620813383 & 20.95 & 2.34 \\
        & 59000.0632999980 & 37.16 & 10.25 \\
        & 59000.0638772579 & 19.21 & 1.12 \\
        & 59000.0738830964 & 41.32 & 11.88 \\
        & 59000.0770900960 & 107.81 & 8.84 \\
        & 59000.0792708558 & 93.70& 10.25 \\
        & 59000.0805536557 & 67.76 & 4.90 \\
        & 59000.0911367554 & 42.02 & 5.68 \\
        & 59000.0954982755 & 86.67 & 7.63 \\
        & 59000.0961396755 & 59.39 & 11.88 \\
        & 59000.0962038155 & 32.21 & 15.96 \\
        & 59000.0981280156 & 37.15 & 11.88 \\
        & 59000.1039006160 & 51.58 & 7.63 \\
        & 59000.1123029571 & 68.32& 7.63 \\
        & 59000.1149968375 & 33.48 & 6.58 \\
        & 59000.1208335787 & 71.03 & 7.63 \\
        & 59000.1222446590 & 67.53 & 7.63 \\
        & 59000.1267344600 & 73.46& 10.25 \\
        & 59000.1268627401 & 66.17 & 8.84 \\
        & 59000.1357140626 & 60.51 & 6.58 \\
        & 59000.1378948233 & 65.36 & 24.85 \\
        & 59000.1379589633 & 17.06 & 2.34 \\
        & 59001.0568292655 & 51.10 & 10.25 \\
        & 59001.0627301540 & 35.66 & 4.90 \\
        & 59001.0676048012 & 80.90 & 10.25 \\
        & 59001.0747884921 & 74.62& 8.84 \\
        & 59001.0779313570 & 51.85 & 10.25 \\
        & 59001.0820254615 & 41.06 & 8.84 \\
        & 59001.0881937735 & 52.96 & 8.84 \\
        & 59001.0895407157 & 94.51 & 10.25 \\
        & 59001.0897331360 & 32.97 & 11.88 \\
        & 59001.0931967018 & 93.64 & 8.84 \\
        & 59001.0933249820 & 45.31 & 10.25 \\
        & 59001.0935815425 & 39.02 & 8.84 \\
        & 59001.0942229435 & 67.29 & 4.90 \\
        & 59001.1021121771 & 45.70 & 8.84 \\
        & 59001.1026252980 & 40.24 & 1.29 \\
        & 59001.1093600099 & 25.66 & 2.02 \\
        & 59001.1097448506 & 109.78 & 4.90 \\
        & 59001.1148119197 & 41.30 & 10.25 \\
        & 59001.1157098814 & 70.42 & 10.25 \\
        & 59001.1158381616 & 57.13 & 10.25 \\
        & 59001.1173972215 & 61.93 & 8.84 \\
        & 59001.1239198167 & 45.63 & 13.77 \\
        & 59001.1251384790 & 28.61 & 11.88 \\
        & 59001.1252667592 & 49.64 & 11.88 \\
        & 59001.1306545296 & 35.09 & 13.77 \\
        & 59001.1355933193 & 58.60 & 7.63 \\
        & 59001.1379023638 & 68.37 & 10.25 \\
\end{tabular}
\end{table}

\newpage
\begin{table}
\addtocounter{table}{-1}
\caption{, continued}
\centering
\begin{tabular}{l c c c}
\hline\hline
Telescope  &  Topocentric MJDs & S/N & Width (ms) \\ [0.5ex] 
\hline
MkII    & 59009.0348197677 & 47.49 & 4.90 \\
        & 59009.0361025867 & 52.82 & 10.25 \\
        & 59009.0368722781 & 16.63 & 3.64 \\
        & 59009.0381550971 & 56.19 & 13.77 \\
        & 59009.0387037037 & 90.17 & 5.68 \\
        & 59009.0502135961 & 22.73 & 13.77 \\
        & 59009.0532282211 & 61.78 & 8.84 \\
        & 59009.0702897166 & 57.76 & 5.68 \\
        & 59009.0736250468 & 20.62 & 3.14 \\
        & 59009.1101854022 & 50.14 & 5.68 \\
        & 59009.1180747432 & 43.66 & 11.88 \\
        & 59015.0571091614 & 34.57 & 8.84 \\
        & 59015.0586485605 & 29.98 & 8.84 \\
        & 59015.0730162863 & 63.94 & 8.84 \\
        & 59015.0749405355 & 17.94 & 10.25 \\
        & 59015.0750046771 & 61.14 & 7.63 \\
        & 59015.0780834757 & 52.63 & 5.68 \\
        & 59015.0807132829 & 27.53 & 4.22 \\
        & 59015.0808415662 & 56.80 & 7.63 \\
        & 59015.0846259232 & 52.11 & 4.22 \\
        & 59015.0848824896 & 20.13 & 8.84 \\
        & 59015.0852031978 & 69.51 & 6.58 \\
        & 59015.0862810631 & 39.22& 8.84 \\
        & 59015.0936057531 & 30.75 & 6.58 \\
        & 59015.1014310341 & 33.26 & 7.63 \\
        & 59020.0113011637 & 49.70 & 7.63 \\
        & 59020.0121350120 & 80.24 & 7.63 \\
        & 59020.0197679314 & 49.98 & 15.96 \\
        & 59020.0218204812 & 54.77 & 11.88 \\
        & 59020.0363807573 & 89.90 & 6.58 \\
        & 59054.9119285904 & 89.25 & 4.22 \\
        & 59054.9295687414 & 126.41 & 6.58 \\
\hline
\end{tabular}
\end{table}

\bsp	
\label{lastpage}
\end{document}